\documentclass[10pt,twocolumn,letterpaper]{article}

\usepackage[pagenumbers]{cvpr}      
\usepackage{graphicx}
\usepackage{amsmath}
\usepackage{gensymb}
\usepackage{amssymb}
\usepackage{booktabs}
\usepackage{textcomp}
\usepackage{csquotes}
\usepackage[pagebackref,colorlinks]{hyperref}
\usepackage{dcolumn}
\usepackage{arydshln}
\usepackage{enumitem}
\usepackage{balance}
\usepackage{combelow}
\usepackage{tabularx}
\usepackage{colortbl}
\usepackage{multirow}
\usepackage{makecell}
\usepackage[noend]{algpseudocode}
\usepackage{algorithmicx,eqparbox,array}

\usepackage{listings}

\usepackage{xcolor,colortbl}
\definecolor{Gray}{gray}{0.9}

\usepackage{arydshln}
\usepackage{xspace}
\makeatletter
\def\adl@drawiv#1#2#3{%
        \hskip.5\tabcolsep
        \xleaders#3{#2.5\@tempdimb #1{1}#2.5\@tempdimb}%
                #2\z@ plus1fil minus1fil\relax
        \hskip.5\tabcolsep}
\newcommand{\cdashlinelr}[1]{%
  \noalign{\vskip\aboverulesep
           \global\let\@dashdrawstore\adl@draw
           \global\let\adl@draw\adl@drawiv}
  \cdashline{#1}
  \noalign{\global\let\adl@draw\@dashdrawstore
           \vskip\belowrulesep}}
\makeatother
\usepackage[capitalize]{cleveref}
\crefname{section}{Sec.}{Secs.}
\Crefname{section}{Section}{Sections}
\Crefname{table}{Table}{Tables}
\crefname{table}{Tab.}{Tabs.}

\usepackage[nonumberlist,sort=def]{glossaries-extra}
\newglossary[nlg]{notation}{not}{ntn}{}
\setglossarystyle{tree}
\setglossarystyle{superheaderborder}
\makenoidxglossaries
\glsdisablehyper
\newglossaryentry{image}{
    type=notation,
    name=\ensuremath{x},
    description={2D image}
}

\newglossaryentry{imageWidth}{
    type=notation,
    name=\ensuremath{W},
    description={Image width}
}

\newglossaryentry{imageHeight}{
    type=notation,
    name=\ensuremath{H},
    description={Image height}
}

\newglossaryentry{viewWidth}{
    type=notation,
    name=\ensuremath{\tilde W},
    description={2D view width}
}

\newglossaryentry{viewHeight}{
    type=notation,
    name=\ensuremath{\tilde H},
    description={2D view height}
}

\newglossaryentry{textureWidth}{
    type=notation,
    name=\ensuremath{W_T},
    description={Texture image width}
}

\newglossaryentry{textureHeight}{
    type=notation,
    name=\ensuremath{H_T},
    description={Texture image height}
}

\newglossaryentry{viewHeightCoord}{
    type=notation,
    name=\ensuremath{\tilde y},
    description={Y-axis coordinate of a 2D view where lesion is placed}
}

\newglossaryentry{viewWidthCoord}{
    type=notation,
    name=\ensuremath{\tilde x},
    description={X-axis coordinate of a 2D view where lesion is placed}
}

\newglossaryentry{vertices}{
    type=notation,
    name=\ensuremath{V},
    description={Set of mesh vertices}
}

\newglossaryentry{faces}{
    type=notation,
    name=\ensuremath{F},
    description={Set of mesh faces}
}

\newglossaryentry{face}{
    type=notation,
    name=\ensuremath{f},
    description={A mesh face}
}

\newglossaryentry{textureImage}{
    type=notation,
    name=\ensuremath{T},
    description={2D texture image}
}

\newglossaryentry{blendedTextureImage}{
    type=notation,
    name=\ensuremath{T_b},
    description={2D texture image w/ blended skin conditions}
}

\newglossaryentry{textureMask}{
    type=notation,
    name=\ensuremath{T_m},
    description={2D texture mask of blended skin conditions}
}

\newglossaryentry{nonSkinTextureImage}{
    type=notation,
    name=\ensuremath{T_{\mathrm{nonskin}}},
    description={2D texture mask of non-skin regions}
}

\newglossaryentry{uvMap}{
    type=notation,
    name=\ensuremath{U},
    description={Set of UV texture coordinates}
}

\newglossaryentry{segMask}{
    type=notation,
    name=\ensuremath{s},
    description={2D binary segmentation mask}
}

\newglossaryentry{view2d}{
    type=notation,
    name=\ensuremath{\tilde a},
    description={2D view of a 3D mesh}
}

\newglossaryentry{viewDepth2d}{
    type=notation,
    name=\ensuremath{\tilde z},
    description={2D view w/ depth values}
}

\newglossaryentry{viewOriginal2d}{
    type=notation,
    name=\ensuremath{\tilde a_{T}},
    description={2D view w/ original textures}
}

\newglossaryentry{viewPaste2d}{
    type=notation,
    name=\ensuremath{\tilde a_{T_p}},
    description={2D view w/ pasted skin condition}
}

\newglossaryentry{viewDilated2d}{
    type=notation,
    name=\ensuremath{\tilde a_{T_d}},
    description={2D view w/ dilated pasted skin condition}
}

\newglossaryentry{viewBlend2d}{
    type=notation,
    name=\ensuremath{\tilde a_{T_b}},
    description={2D view w/ blended skin condition}
}

\newglossaryentry{viewMask2d}{
    type=notation,
    name=\ensuremath{\tilde a_{T_m}},
    description={2D mask of the skin condition}
}

\newglossaryentry{viewSkinMask2d}{
    type=notation,
    name=\ensuremath{a_{\mathrm{skin}}},
    description={2D mask of the skin}
}

\newglossaryentry{viewNonSkinMask2d}{
    type=notation,
    name=\ensuremath{a_{\mathrm{nonskin}}},
    description={2D mask of the non-skin regions}
}

\newglossaryentry{faces2d}{
    type=notation,
    name=\ensuremath{\tilde f},
    description={Indices of 3D mesh faces in a 2D view}
}

\newglossaryentry{vertices2d}{
    type=notation,
    name=\ensuremath{\tilde y},
    description={3D coordinates of anatomically correspondent template in a 2D view}
}

\newglossaryentry{weightDistance}{
    type=notation,
    name=\ensuremath{w},
    description={Scalar weight of the camera-to-mesh distance}
}

\newglossaryentry{template}{
    type=notation,
    name=\ensuremath{T},
    description={Standardized 3D mesh template}
}

\newglossaryentry{templateVertices}{
    type=notation,
    name=\ensuremath{T_v},
    description={Template's 3D vertices}
}

\newglossaryentry{templateEdges}{
    type=notation,
    name=\ensuremath{T_e},
    description={The edges between template mesh's vertices}
}

\newglossaryentry{templateFaces}{
    type=notation,
    name=\ensuremath{T_f},
    description={Faces of the template mesh}
}

\newglossaryentry{templateImage}{
    type=notation,
    name=\ensuremath{T_x},
    description={Template's texture image}
}

\newglossaryentry{parameters}{
    type=notation,
    name=\ensuremath{\theta},
    description={Parameters of an ML model}
}

\newglossaryentry{learnedParameters}{
    type=notation,
    name=\ensuremath{\theta^*},
    description={Learned parameters of an ML model}
}

\newglossaryentry{imageVertices}{
    type=notation,
    name=\ensuremath{y},
    description={Mesh vertices visible in a 2D image}
}

\newglossaryentry{predictedImageVertices}{
    type=notation,
    name=\ensuremath{\hat{y}},
    description={Predicted mesh vertices visible in a 2D image}
}

\newglossaryentry{mesh}{
    type=notation,
    name=\ensuremath{M},
    description={3D mesh}
}

\newglossaryentry{meshVertices}{
    type=notation,
    name=\ensuremath{M_v},
    description={3D vertices of a mesh}
}

\newglossaryentry{meshFaces}{
    type=notation,
    name=\ensuremath{M_f},
    description={Faces of a mesh}
}

\newglossaryentry{meshEdges}{
    type=notation,
    name=\ensuremath{M_e},
    description={The edges between mesh's vertices}
}

\newglossaryentry{matchedMesh}{
    type=notation,
    name=\ensuremath{\tilde M},
    description={A mesh that maintains the original structure but has vertex indices that correspond to the template mesh vertices}
}

\newglossaryentry{cameraParameters}{
    type=notation,
    name=\ensuremath{\kappa},
    description={Camera parameters for a 2D view}
}

\newglossaryentry{lightingParameters}{
    type=notation,
    name=\ensuremath{L},
    description={Lighting parameters for rendering}
}

\newglossaryentry{materialParameters}{
    type=notation,
    name=\ensuremath{\mathcal{M}},
    description={Material parameters for rendering}
}

\usepackage{amsmath}
\DeclareMathOperator*{\argmin}{argmin}

\usepackage{xcolor}
\definecolor{background}{RGB}{37,37,37}
\definecolor{skin}{RGB}{77,154,74}
\definecolor{lesion}{RGB}{222,45,38}
\definecolor{real}{RGB}{77,154,74}
\definecolor{synth}{RGB}{222,45,38}

\definecolor{codegreen}{rgb}{0,0.6,0}
\definecolor{codegray}{rgb}{0.5,0.5,0.5}
\definecolor{codepurple}{rgb}{0.58,0,0.82}
\definecolor{backcolour}{rgb}{0.95,0.95,0.92}
\definecolor{cb_lightblue}{RGB}{0,114,178}
\definecolor{cb_orange}{RGB}{213,114,0}
\definecolor{cb_yellow}{RGB}{230,159,0}
\definecolor{cb_darkblue}{RGB}{0,114,178}
\definecolor{cb_vermillion}{RGB}{204,121,167}
\definecolor{cb_blueishgreen}{RGB}{0,158,115}
\definecolor{cb1}{HTML}{0173b2}
\definecolor{cb2}{HTML}{de8f05}
\definecolor{cb3}{HTML}{029e73}
\lstdefinestyle{mystyle}{
  backgroundcolor=\color{backcolour},
  commentstyle=\color{codegreen},
  keywordstyle=\color{magenta},
  numberstyle=\tiny\color{codegray},
  stringstyle=\color{codepurple},
  basicstyle=\ttfamily\footnotesize,
  breakatwhitespace=false,
  breaklines=true,
  captionpos=b,
  keepspaces=true,
  numbers=none,
  frame=shadowbox,
  numbersep=5pt,
  showspaces=false,
  showstringspaces=false,
  showtabs=false,
  tabsize=4
}

\definecolor{bg}{HTML}{363A4F}
\newcommand{\method}{\textit{DermSynth3D}} %
\newcommand{\figref}{Figure~\ref}
\newcommand{\secref}{Section~\ref}
\newcommand{\tabref}{Table~\ref}
\newcommand{\eqnref}{Eq.~\ref}

\usepackage[capitalize]{cleveref}
\crefname{section}{Sec.}{Secs.}
\Crefname{section}{Section}{Sections}
\Crefname{table}{Table}{Tables}
\crefname{table}{Tab.}{Tabs.}

\usepackage[fitting]{tcolorbox}
\newcommand*{\affaddr}[1]{#1}

\newcommand*{\email}[1]{\small\texttt{#1}}
\begin{document}
\raggedbottom
\newcommand{\jer}[1]{\textcolor{black}{#1}}
\newcommand{\ash}[1]{\textcolor{black}{#1}}
\newcommand{\ari}[1]{\textcolor{black}{#1}}
\newcommand{\ka}[1]{\textcolor{black}{#1}}
\title{\method{}: Synthesis of in-the-wild Annotated Dermatology Images}%

\author{%
  Ashish Sinha$^1$\thanks{Authors contributed equally (joint first authors)}, Jeremy Kawahara$^1$\footnotemark[1], Arezou Pakzad$^1$\footnotemark[1], Kumar Abhishek$^1$, Matthieu Ruthven$^2$,\\Enjie Ghorbel$^{2,3}$, Anis Kacem$^2$, Djamila Aouada$^2$, and Ghassan Hamarneh$^1$\thanks{Corresponding author}\\ \\
\affaddr{$^1$Medical Image Analysis Lab, School of Computing Science, Simon Fraser University, Canada}\\
\email{\{ashish\_sinha, jkawahar, arezou\_pakzad, kabhishe, hamarneh\}@sfu.ca}\\
\affaddr{$^2$Computer Vision, Imaging \& Machine Intelligence Research Group, Interdisciplinary\\Centre for Security, Reliability and Trust (SnT), University of Luxembourg, Luxembourg}\\
\affaddr{$^3$Cristal Laboratory, National School of Computer Sciences, University of Manouba, 2010, Tunisia}\\
\email{\{firstname.lastname\}@uni.lu}\\
}

\maketitle

\begin{abstract}

In recent years, deep learning (DL) has shown great potential in the field of dermatological image analysis.
However, existing datasets in this domain have significant limitations, including a small number of image samples, limited disease conditions, insufficient annotations, and non-standardized image acquisitions.
To address these shortcomings, we propose a novel framework called \method{}.
\method{} blends skin disease patterns onto 3D textured meshes of human subjects using a differentiable renderer and generates 2D images from various camera viewpoints under chosen lighting conditions in diverse background scenes.
Our method adheres to top-down rules that constrain the blending and rendering process to create 2D images with skin conditions that mimic \emph{in-the-wild} acquisitions, ensuring more meaningful results.
The framework generates photo-realistic 2D dermoscopy images and the corresponding dense annotations for semantic segmentation of the skin, skin conditions, body parts, bounding boxes around lesions, depth maps, and other 3D scene parameters, such as camera position and lighting conditions.
\method{} allows for the creation of custom datasets for various dermatology tasks.
We demonstrate the effectiveness of data generated using \method{} by training DL models on synthetic data and evaluating them on various dermatology tasks using real 2D dermatological images.
We make our code publicly available at \url{https://github.com/sfu-mial/DermSynth3D}.
\end{abstract}

\section{Introduction}

\begin{figure*}[htb]
    \centering
    \includegraphics[width=\linewidth]{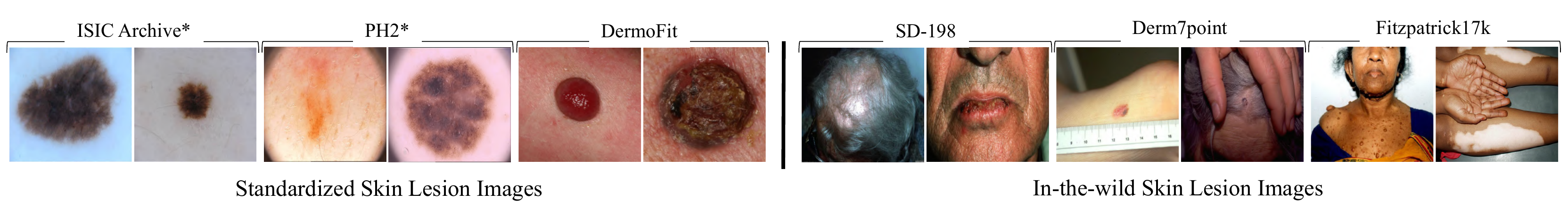}
    \caption{Standardized vs in-the-wild skin lesion images ($*$: dermoscopy, all others: clinical).
    }
    \label{fig:standard-vs-in-the-wild}
\end{figure*}

The diagnosis and analysis of skin conditions are an enormous burden on the healthcare system, with \emph{at least $3000$} distinct skin diseases identified~\cite{Bickers2006} so far.
Both human dermatologists and sophisticated computerized approaches struggle to address this complex task of analyzing skin conditions.
Computerized analysis of skin diseases often rely on 2D colored images, with significant research efforts devoted to analysis of conditions within clinical~\cite{Li2021} and dermoscopy images~\cite{Celebi2019}.
While clinical images can capture a variety of skin conditions using a common digital camera, dermoscopy images offer a more standardized acquisition using a dermatoscope, which captures a highly magnified image of the lesion with details imperceptible to the naked eye.

\emph{Dermoscopy} images generally focus on the analysis of a single lesion, with large scale annotated dermoscopy datasets now available for public use~\cite{Tschandl2018,combalia2019bcn20000,rotemberg2021patient}.
While dermoscopy has been shown to improve the diagnostic ability of trained specialists, the field-of-view of a dermoscopy image is generally limited to a localized patch of skin on the body (e.g., a mole).
In contrast, clinical images vary considerably in their acquisition protocols, ranging from a closeup view focused on a single lesion, to a view that captures a significant portion of the body (\figref{fig:standard-vs-in-the-wild}).
The contextual information in large-scale clinical images of skin lesions may provide valuable cues regarding the underlying disease that may not be present in dermoscopic images alone \cite{rotemberg2021patient,birkenfeld2020computer}.

\emph{Clinical} images exhibit considerable variability across datasets. For example, the public DermoFit Image Library dataset~\cite{Edinburgh, Ballerini2013} contains 1300 clinical images and manual lesion segmentations from 10 types of skin conditions.
These are high-quality images acquired under standardized conditions.
In contrast, other clinical datasets, such as SD-198~\cite{Sun2016}, SD-260~\cite{yang2019self}, or Fitzpatrick17K~\cite{Groh2021}, contain hundreds of types of skin disorders and are much less standardized, exhibiting a high variability in camera position relative to the lesion, resulting in dramatic changes in the field-of-view.
We use the term ``in-the-wild clinical dataset" to describe these types of image collections, where the camera position, field-of-view, and background, are inconsistent.

In-the-wild clinical images are often used to train a classification model~\cite{Sun2016,kawahara2018seven,Groh2021,wu2022fairprune,daneshjou2023skincon}, where the entire image is taken as input, and the model is trained to produce a label (e.g., class of skin disorder). However, there are several important dermatological tasks apart from classification of skin disorders, such as lesion segmentation~\cite{mirikharaji2022survey,hasan2023survey}, lesion tracking~\cite{young2021role, fried2020technological, sondermann2019tracking}, lesion management~\cite{abhishek2021predicting}, and skin tone prediction~\cite{kinyanjui2019estimating}.
As an example, \cite{Wang2020} motivated their release of a public wound segmentation dataset of 2D clinical images by noting that wound segmentation may help automate the process of measuring the wound area to monitor healing and determine therapies.
In addition, \cite{gholami2017segmentation} showed that chronic wound bioprinting based on image segmentation can help facilitate wound treatments.  \cite{Groh2021} created a public dataset of 2D clinical images with skin disorder and Fitzpatrick skin tone labels~\cite{fitzpatrick1975soleil,archderm1988} and noted the need to segment pixels containing healthy skin when applying automated methods to estimate the skin tones of the imaged subjects. Other works~\cite{Lee2005,Mirzaalian2016,TotalBodyPhotography2018,zhao2021skin3d} have motivated the importance of considering multiple lesions over a widely imaged area, as opposed to focusing on a single lesion, noting that the presence of multiple nevi (moles) is an important indicator for melanoma~\cite{Gandini2005}.

One approach to curate the necessary data is to synthesize images with their corresponding annotations, which has shown success in other domains, both medical and non-medical. For example, for non-medical applications, image synthesis with annotations has been used in face analysis~\cite{Wood2021} and indoor scene segmentation~\cite{McCormac2017}.
For a more comprehensive review of image synthesis, particularly using generative adversarial network (GAN) models~\cite{goodfellow2014generative,wang2021generative}, we direct the interested readers to the survey by \cite{shamsolmoali2021image}. Since medical image datasets tend to be small~\cite{kohli2017medical,curiel2019artificial,asgari2021deep}, synthesis for medical image analysis applications has also gained popularity in recent years to generate ground truth-annotated images, including but not limited to MRI~\cite{chartsias2017multimodal,dar2019image}, CT~\cite{nie2017medical,chuquicusma2018fool}, PET~\cite{bi2017synthesis,wang20183d}, and ultrasound~\cite{tom2018simulating,liang2022sketch}. For a more in-depth review of the use of GANs and image synthesis in medical imaging, we refer the interested readers to comprehensive surveys by \cite{yi2019generative}, \cite{kazeminia2020gans}, \cite{wang2021review}, \cite{rawat2023gans}, and \cite{Yang2023-gw}.

Similarly, for skin image analysis, there have been several works towards the synthesis of skin lesion images. The first two works to explore skin lesion image synthesis used a variety of noise-based GANs \cite{baur2018generating} and conditioned the output on the diagnostic category \cite{bissoto2018skin}.
\cite{abhishek2019mask2lesion} then proposed a GAN-based framework to generate skin lesion images constrained to binary lesion segmentation masks, while \cite{pollastri2020augmenting} used GANs to generate both skin lesion images as well as the corresponding binary segmentation masks.
For a more detailed review of the literature on deep learning-based synthetic data generation for skin lesion images, we refer interested readers to the comprehensive survey by \cite{mirikharaji2022survey}.

While there are numerous publicly available 2D dermatological image datasets \cite{mirikharaji2022survey}, existing ``in-the-wild" clinical datasets have limitations in creating semantically rich ground truth (GT) labels that can be used for the diverse range of dermatological tasks discussed earlier. Consequently, compared to dermoscopic images' synthesis, there is considerably less research in the synthetic data generation of clinical images. \cite{Li2017} proposed to synthesize 2D data by blending small lesions onto a larger 2D image of the torso, which allowed them to create training data for a neural network that detects lesions' masks across a large region of the body. \cite{Dai2021} proposed to generate burn images with automatic annotations.
They used a Style-GAN~\cite{karras2019style} to synthesize burn wounds, blended the generated burns with textures from a 3D human avatar, and generated a 2D training dataset through sampling from different 2D views of the 3D avatar with the synthetic burns. Both approaches motivated their use of synthetic data by noting the difficulties in collecting appropriate real labeled training data that is specific to their dermatological task.

Our proposed work is similar to that of \cite{Dai2021} in that we follow a similar pipeline where 2D images of the skin disorder are blended onto the 3D textured meshes and used to create a large-scale 2D dataset with corresponding annotations.
However, we extend this framework by incorporating a deep blending approach to blend lesions \emph{across seams} in 2D rendered views.
Additionally, we broaden the scope of this work by including a diverse range of skin tones and background scenes, enabling us to generate semantically rich and meaningful labels for 2D \emph{in-the-wild} clinical images that can be used for a variety of dermatological tasks, as opposed to just one.

\begin{center}
    \captionsetup{type=table}
    \captionof{table}{Summary of Notations}
    \vspace{-1em}
    \printnoidxglossaries
    \label{table:glossary}
\end{center}

\ash{
Furthermore, the annotated data generated by \method{} in the form of semantic segmentation masks, depth maps, and 3D scene parameters, can be used to train machine learning models for a variety of medical tasks that may benefit clinical practice.
For instance, the scene parameters may be used to train models for reconstruction and visualization of 3D anatomical organs, longitudinal tracking of lesions, illumination, and skin tone estimation for consistent imaging and tracking.
The surgeons can use these reconstructed 3D models for pre-operative planning, allowing them to better visualize the patient's anatomy and anticipate potential challenges.
Longitudinal tracking of lesions can help the doctors in measuring the progress of diseases, evaluating the effectiveness of treatments and administer better-suited treatments.
The measurement bias introduced across time due to change in background and lighting conditions can be further corrected by training deep models to accurately estimate the illumination, skin-tone and camera parameters.
}
To facilitate future extensions to our framework, we have made our code base highly modular and publicly available.

\subsection{Contributions}
Despite the availability of numerous skin image datasets
(e.g., ~\cite{Ballerini2013,Tschandl2018,kawahara2018seven,Wang2020,Groh2021,wen2022characteristics,daneshjou2023skincon}),
there is a lack of a \emph{large-scale} skin-image dataset that can be applied to a variety of skin analysis tasks, especially in an \emph{in-the-wild} clinical setting.
Moreover, existing datasets are limited in their scope and are often task-specific, requiring extensive additional annotation for generalizing them to other dermatological applications.

\vspace{1em}
\begin{lstlisting}[caption={A minimal example of the code showing the usage of our proposed pipeline \method{}, which illustrates the process of selecting a location to place the lesion on the mesh, followed by blending the lesion with the texture image, and finally rendering 2D synthetic views with corresponding labels. Infinite variations such as lighting, viewpoints, lesions, \emph{etc.} can be specified in the configurations.}, label={listing:code_snippet}]
from dermsynth3d import (SelectAndPaste,
                         BlendLesions,
                         Generate2DViews)

# Load settings stored in YAML file, such as:
# file paths, number of lesions to blend,
# scene parameters for the renderer, etc.
config = (...)

select_locations = SelectAndPaste(config)
select_locations.paste_on_locations()

blender = BlendLesions(config)
blender.blend_lesions()

renderer = Generate2DViews(config)
renderer.synthesize_views()
\end{lstlisting}

To address this gap, we present \method{}, a computational pipeline along with an open-source software library, for generating synthetic 2D skin image datasets using 3D human body meshes blended with skin disorders from clinical images.
Our approach uses a differentiable renderer to blend the skin lesions within the texture image of the 3D human body and generates 2D views along with corresponding annotations, including semantic segmentation masks for skin conditions, healthy skin, non-skin regions, and anatomical regions.
Furthermore, we demonstrate the utility of the synthesized data by using it to train machine learning models and evaluating them on real-world dermatological images, showcasing that the \method{}-trained model learns to generalize to a variety of dermatological tasks.
Additionally, the open-source and modular design of our framework offers opportunities for researchers in the community to experiment and choose from a range of 2D skin disorders, renderers, 3D scans, and various other scene parameters.
We present a simplified code snippet in Listing \ref{listing:code_snippet} that exemplifies the modular implementation of our proposed framework and emphasizes its user-friendliness and ease of use.






\section{Methods}

\begin{figure}[h]
    \centering
    \includegraphics[width=\linewidth]{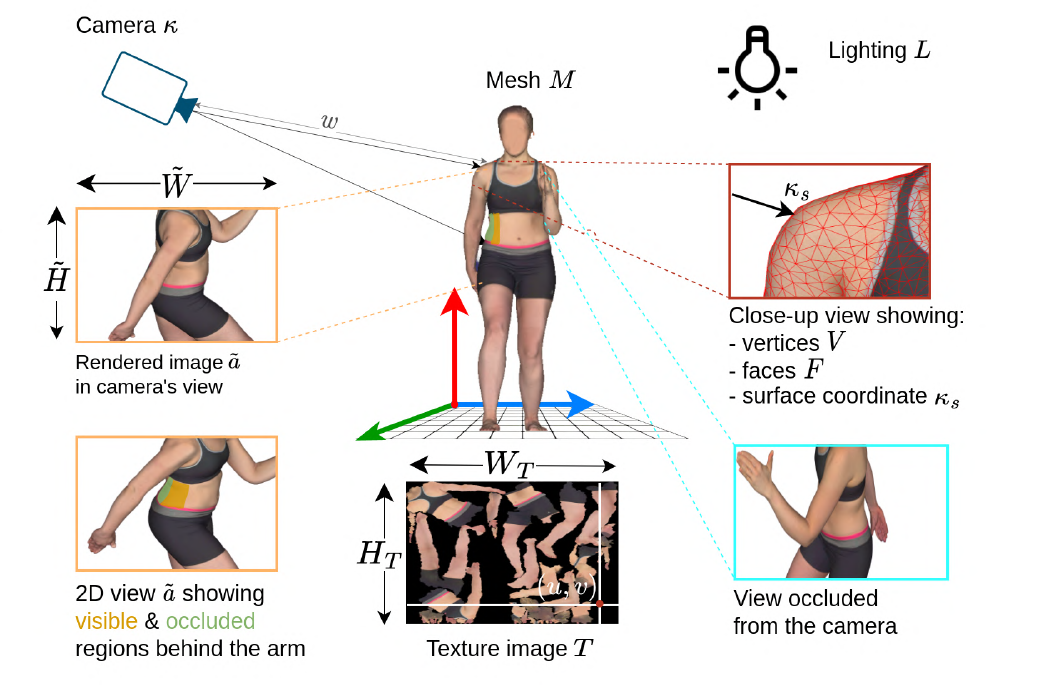}
    \caption{A figure depicting the essential notations illustrating the conditions for camera positioning, which require that the camera is positioned outside the mesh and that the mesh does not obstruct the light rays.}
    \label{fig:notations}
    \vspace{-1em}
\end{figure}

Our proposed \method{} framework automates the process of blending skin disease regions from 2D images onto 3D texture meshes, while allowing for control over lighting and material parameters, from appropriate camera viewpoints, and renders the resulting 2D image and the corresponding ground truth annotations.
\figref{fig:pipeline} shows our proposed framework.
Here, we describe the rendering of a single image of a single mesh augmented with a single lesion.
However, as these steps are automated, large-scale, multi-lesion datasets can be easily generated. We provide a summary of the mathematical notations used in this paper in Table~\ref{table:glossary} and \figref{fig:notations}.

We define a 2D clinical image $\gls{image} \in \mathbb{R}^{\gls{imageWidth} \times \gls{imageHeight} \times 3}$ as an RGB image with width $W$ and height $H$ that shows a skin condition, and a corresponding binary segmentation mask~$\gls{segMask} \in \{0, 1\}^{\gls{imageWidth} \times \gls{imageHeight}}$ where pixels with a non-zero value indicate the diseased region (as shown in ``2D Lesions" in \figref{fig:pipeline}).
We define a 3D avatar of a human subject as a mesh~\gls{mesh} composed of vertices~\gls{vertices}, faces~\gls{faces}, and a UV map~\gls{uvMap}, where the vertices and the faces determine the geometry of the mesh and the UV map determines the mapping between the geometry and a 2D texture image~$\gls{textureImage} \in \mathbb{R}^{W_T \times H_T \times 3}$ that contains pixels representing the surface of the skin.
Our goal is to transfer the skin condition within \gls{image} onto a location on the texture image \gls{textureImage} of the 3D mesh~\gls{mesh}. We approach this problem through an image-blending approach, where given a 2D binary segmentation mask~\gls{segMask} indicating the skin condition within \gls{image} and a target location on the mesh, we blend the diseased region within the mesh's texture image~\gls{textureImage}.

\subsection{Image Synthesis via Differentiable Rendering}

\begin{figure*}[htb]
    \centering
    \includegraphics[width=\textwidth]{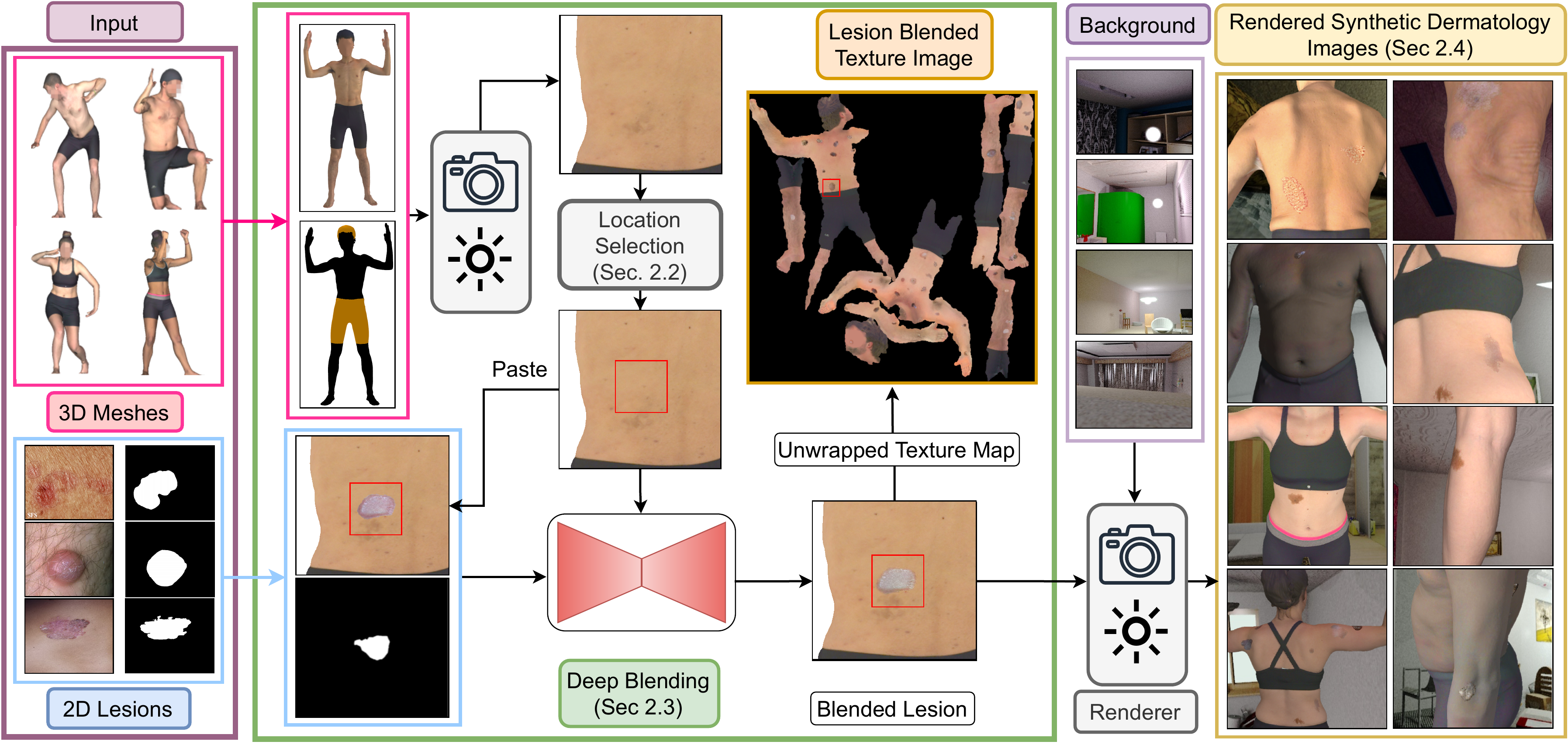}
    \caption{\ash{Overview of our proposed framework \method. The pipeline takes 2D segmented skin conditions and texture image of a 3D mesh as input, and blends the skin condition onto it to produce a lesion blended texture map.
    After blending, 2D views of the mesh are rendered from various camera viewpoints, under different lighting conditions, and combined with background images to create a synthetic dermatology dataset.}}
    \label{fig:pipeline}
\end{figure*}

A straightforward approach to blend a 2D image with a 3D model would be to blend \gls{image} directly with \gls{textureImage}. However, as can be seen in the “texture image'' in \figref{fig:pipeline}, this is challenging as the 2D texture image splits the human body based on seams and maps to 2D regions that are not semantically localized in 3D, which can pose a challenge for larger skin conditions that may span across seams.
To address this, our proposed approach blends skin patterns on a 2D view \gls{view2d} of a 3D mesh rendered using a differentiable renderer, from various camera viewpoints under chosen lighting conditions.
We also impose constraints to avoid blending skin patterns at unsuitable locations such as across disjoint anatomy.
We describe each component of our proposed approach in detail in the following sections.

\ash{We employ PyTorch3D~\cite{Ravi2020} as a differentiable renderer $R(\cdot)$ in our pipeline to render 2D images from meshes, owing to the wide adoption of PyTorch3D in state-of-the-art works employing differntiable rendering techniques.
$R(\cdot)$ is composed of two rendering components, a rasterizer and a shader.
The rasterizer identifies visible faces and computes fragment data, \ie, face indices per pixel, barycentric coordinates, and distances from the camera to the surface of the 3D object.
The shader module calculates the final pixel values by incorporating various factors, including lighting conditions, material properties, and the fragment data computed during rasterization.
In essence, the shader imparts colors and shading to each pixel, culminating in a visually coherent representation of the 3D scene on the 2D image.
$R(\cdot)$ unifies rasterization and shading into a single differentiable renderer that takes 3D object and scene parameters as input and outputs a 2D image and fragment data.
}
\ash{More formally,}
given the mesh \gls{mesh}, texture image \gls{textureImage}, intrinsic and extrinsic camera parameters \gls{cameraParameters}, lighting parameters \gls{lightingParameters}, material parameters \gls{materialParameters}, and a rendered view width $\gls{viewWidth}$ and height $\gls{viewHeight}$, we render a 2D view \gls{view2d} of the 3D mesh:
\begin{equation}
    \gls{view2d}, \gls{faces2d}, \tilde z  = R(\gls{mesh}, \gls{textureImage}; \gls{cameraParameters}, \gls{lightingParameters}, \gls{materialParameters}, \gls{viewWidth}, \gls{viewHeight})
    \label{eq:renderer}
\end{equation}
where the rendered 2D view $\gls{view2d} \in \mathbb{R}^{\gls{viewWidth} \times \gls{viewHeight} \times 3}$ has the given width and height; $\gls{faces2d} \in \mathbb{Z}^{\gls{viewWidth} \times \gls{viewHeight}}$ indicates the indices of the faces of the mesh that are visible within \gls{view2d}; $\gls{viewDepth2d} \in \mathbb{Z}^{\gls{viewWidth} \times \gls{viewHeight}}$ indicates the depth of the mesh with respect to the camera position for each pixel in \gls{view2d}; and, $R(\cdot)$ is the differentiable rendering function. The camera parameters \gls{cameraParameters} control
where on the body the skin condition is to be blended
and \gls{lightingParameters} and \gls{materialParameters} describe the lighting and material parameters respectively which jointly control the visual appearance.

\subsection{Determination of Skin Condition Location on the Mesh}
\label{sec:blend_locations}
While skin conditions can manually be placed on different potential locations on the mesh,
we {enforce} our automated approach to place skin conditions only at suitable locations.
We define the following criteria for a suitable location:
the region where a lesion can be placed on the mesh, should; (1) not overlap with clothes or the hair on the head, (2) not overlap with the background, (3) have minimal depth changes, preventing blending lesions across disjoint anatomy.
Specifically, as shown in \figref{fig:notations} (the 2D view $\tilde{a}$) for instance, we must avoid blending a lesion that extends across and covers both the right arm and the torso behind the right arm.
Also, when blending multiple skin conditions, we ensure that skin conditions do not overlap.
Accordingly, to ensure that the human anatomy remains in the camera's field of view, we constrain the camera's viewpoint to point to a specific coordinate $\gls{cameraParameters}_{s}$ on the surface of the mesh, referred to as the surface coordinate.
To determine the camera's position in the world space, we compute the weighted sum of the surface coordinate and its normal vector, where the weight is controlled by the parameter \gls{weightDistance}.
We express this operation as:
\begin{equation}
    \gls{cameraParameters}_{p} = \gls{cameraParameters}_{s} + n(\gls{cameraParameters}_{s}) * \gls{weightDistance}
    \label{eq:camera_position}
\end{equation}
where $\gls{cameraParameters}_{p} \in \mathbb{R}^{3}$ is the camera position; $\gls{cameraParameters}_{s} \in \mathbb{R}^{3}$ is the surface coordinate; and $n(\gls{cameraParameters}_{s})$ is the normal at the surface coordinate $\gls{cameraParameters}_{s}$.
The scalar weight \gls{weightDistance} controls the distance from the camera to the surface of the mesh, where a larger weight places the camera further from the surface, i.e., captures a larger field of view.
Sampling the weights \gls{weightDistance} and surface coordinates $\gls{cameraParameters}_{s}$ results in a range of views; however, many views are unsuitable to place the skin condition.

Thus, first, we check the suitability of placing a scaled clinical image \gls{image} and lesion mask \gls{segMask} (scaled as \gls{image} and \gls{view2d} can be different sizes) at the center of the rendered view, by first composing an image $a_x \in \mathbb{R}^{\gls{viewWidth} \times \gls{viewHeight} \times 3}$ (showing the lesion on the rendered view) and a corresponding lesion mask $a_s \in \mathbb{R}^{\gls{viewWidth} \times \gls{viewHeight}}$.
We then check if the region  $a_s$, where the skin disorder was placed meets the aforementioned criteria.
For ensuring minimal depth changes and avoiding lesion overlap with the background,
we rely on the depth \gls{viewDepth2d} from the renderer (\eqnref{eq:renderer}) to avoid local regions that have a high depth change or are outside the mesh.
Moreover, to avoid lesion overlap with non-skin regions,
we use manual annotations (\secref{sec:3dbodytex}) of non-skin regions on the texture image to distinguish skin from non-skin regions. Further details are supplied in \ref{sup:sec:methods}.

\begin{figure*}[htb]
    \centering
    \includegraphics[width=0.8\linewidth]{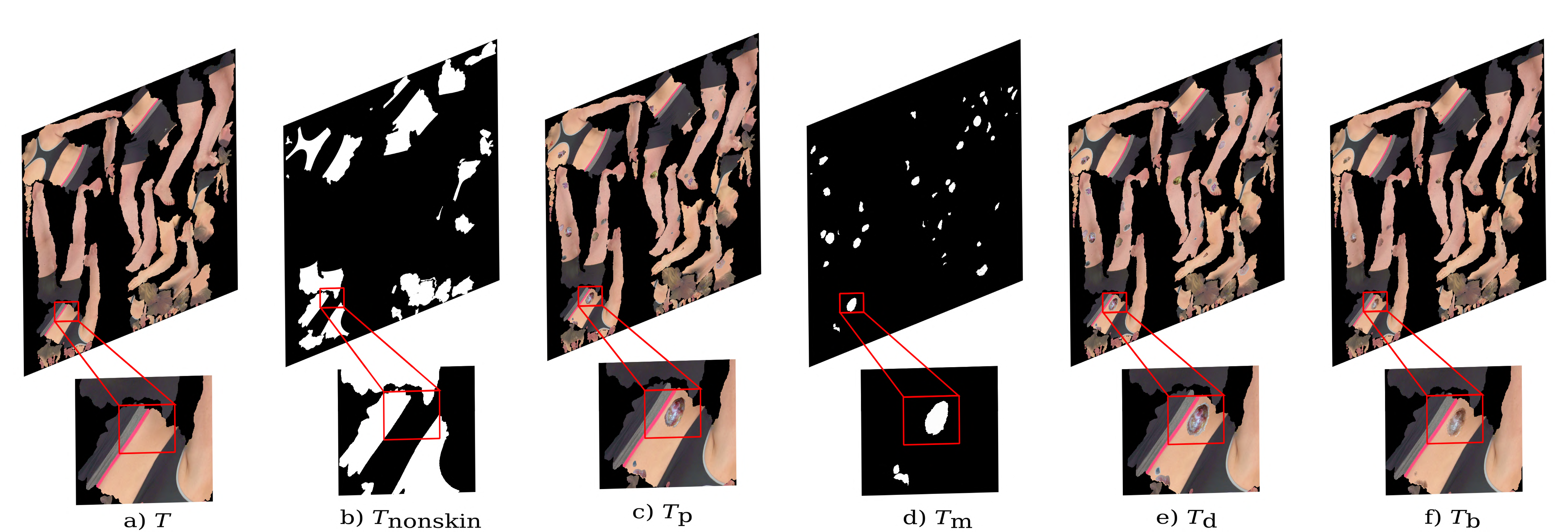}
    \caption{The different variants of texture images produced by \method{} with a corresponding close view of skin -conditions.
    Left to right: Original texture image \gls{textureImage}, Binary mask indicating the non-skin regions \gls{nonSkinTextureImage}, texture image with the ``pasted" skin  conditions $T_p$, a mask of the texture image showing the localizations of the pasted skin -conditions \gls{textureMask}, a texture image $T_d$ created by dilating the masked lesion $s$, a texture image $T_b$ with the ``blended" skin conditions.}
    \label{fig:all_tex_maps}
\end{figure*}

\subsection{Blending Skin Conditions into a Mesh's Texture Map}
\label{sec:blending}
Skin conditions can appear in different locations on the body and they can be captured in various views in real-world clinical settings.
Thus, in order to efficiently synthesize realistic ``in-the-wild" clinical images from different 3D viewpoints, %
our approach blends the skin disorder within the mesh's texture image, which allows
the framework %
to render the blended skin disorder from a variety of views. %
In order for the blending to be robust to different viewpoints, we perturb the camera position during the blending process using a pasted texture image $T_p$ containing the original skin conditions ``pasted" within the texture image. Specifically,
 given the regions that capture a skin condition within the view $a_x$ and the corresponding mask $a_s$, we map the masked pixels containing the skin disorder to the pixels within the texture image in order to ``paste" the original segmented skin condition onto the texture image \gls{textureImage}.
This mapping is achieved by determining the UV or texture coordinates of a vertex on the surface of the mesh corresponding to its Cartesian coordinate in the texture image.

Given a texture image (\figref{fig:all_tex_maps}-a),  a skin mask (\figref{fig:all_tex_maps}-b),  and an image of a skin condition and its mask (\figref{fig:pipeline}- 2D Lesion, lower-left), our goal is to update the texture image to include a skin condition. We denote this updated texture image with the pasted skin conditions as $T_p$ (\figref{fig:all_tex_maps}-c).
Following the same procedure, we create a texture image mask \gls{textureMask} (\figref{fig:all_tex_maps}-d) that localizes the pasted skin condition, where instead of assigning the pixel value, we assign a unique integer to identify each lesion.
We create an additional texture image $T_d$ (\figref{fig:all_tex_maps}-e), which is based on dilating the masked lesion \gls{segMask} to include the surrounding skin.
We define \gls{blendedTextureImage} (\figref{fig:all_tex_maps}-f) as the texture image that will be optimized during blending and is initialized with $T_p$.
By replacing the input texture image \gls{textureImage} in \eqnref{eq:renderer} with $T_p$, \gls{textureMask}, $T_d$, and \gls{blendedTextureImage}, we can render 2D views of the original unmodified textures (\gls{viewOriginal2d}), the pasted skin condition (\gls{viewPaste2d}), a mask of the pasted skin condition (\gls{viewMask2d}), a 2D view of the dilated skin condition (\gls{viewDilated2d}), and a 2D view of the blended skin condition (\gls{viewBlend2d}), while keeping the camera parameters unchanged.

To create a 2D blended image, we follow the deep image blending approach by \cite{Zhang2020}, where an iterative optimization, minimizes a blending loss function between a foreground object cropped from the source image and the target image which the selected object would be blended onto.
We use this approach to create the 2D blended image $b$ by combining the non-dilated masked pixels \gls{viewMask2d} of the lesion in the blended view \gls{viewBlend2d} with the non-masked pixels containing original textures \gls{viewOriginal2d},
\begin{equation}
    b = \gls{view2d}_{T_m} \odot \gls{view2d}_{T_b} + (1-\gls{view2d}_{T_m}) \odot \gls{view2d}_{T},
\end{equation}
where $\odot$ represents element-wise (Hadamard) multiplication.
This masking causes only the pixels within the masked region to be modified while preserving the original non-masked regions.

However, in contrast to \cite{Zhang2020}, instead of directly blending a 2D image, we optimize the pixels within a texture image \gls{blendedTextureImage} by  minimizing the blending loss using the 2D view,
\begin{equation}
\gls{blendedTextureImage}^* =
\argmin_{{\gls{textureImage}_b}}
\mathcal{L}
\left(
b,
\gls{viewBlend2d}, \gls{viewOriginal2d}, \gls{viewPaste2d}, \gls{viewDilated2d}, \gls{viewMask2d}
\right),
\label{eq:blend}
\end{equation}
where $\gls{blendedTextureImage}^*$ is the resulting texture image with the blended lesions after optimization; and, $\mathcal{L}(\cdot)$ denotes the blending loss. %

We adopt the content $\mathcal{L}_c$, style $\mathcal{L}_s$, gradient $\mathcal{L}_{\nabla}$, and total variation $\mathcal{L}_{\mathrm{TV}}$ loss functions for image blending as described
by~\cite{Zhang2020},
\begin{equation}
\begin{split}
    \mathcal{L} = &
    \ \lambda_c \mathcal{L}_c(b, \gls{view2d}_{T_p}; \gls{view2d}_{T_m}) \\
    & + \lambda_s \mathcal{L}_s( b , \gls{view2d}_{T}; \gls{view2d}_{T_m}) \\
    & + \lambda_{\nabla} \mathcal{L}_{\nabla}( b, \gls{viewOriginal2d}, \gls{viewDilated2d}; \gls{viewMask2d}) \\
    & + \lambda_{\mathrm{TV}} \mathcal{L}_{\mathrm{TV}}( b; \gls{viewMask2d})\ka{,}
\label{eq:blend_loss}
\end{split}
\end{equation}
\ash{Similar to~\cite{Zhang2020}, we use the VGG-16 network pretrained on ImageNet~\cite{deng2009imagenet} to extract features for computing style and content losses, which are defined as,
}
\ash{
\begin{equation}
    \begin{split}
        \mathcal{L}_c = &
        \parallel \mathcal{F} ( b \odot \gls{view2d}_{T_m}) - \mathcal{F} ( \gls{view2d}_{T_p} \odot \gls{view2d}_{T_m}) \parallel_2 \\
        \mathcal{L}_s = &
        \frac{1}{L} \sum_{l=1}^L \parallel \mathcal{G}_l (b \odot \gls{view2d}_{T_m}) - \mathcal{G}_l (\gls{view2d}_{T} \odot \gls{view2d}_{T_m}) \parallel_2 \\
    \end{split}
\end{equation}
where $L$ is the number of convolutional layers in the VGG-16 network $\mathcal{F}(\cdot)$, $\mathcal{G}_l (\cdot)$ is the Gram matrix computed for the features at the $l^{th}$ layer, and $\parallel \cdot \parallel$ denotes the $L_2$-norm.
$\mathcal{L}_c$ encourages the spatial similarity between the image features of blended and pasted views extracted from $\mathcal{F}$, while $\mathcal{L}_s$ encourages the similarity in style or texture between the rendered views of blended and original texture maps.
In order to promote a seamless boundary around the blending region, we employ the Laplacian filter as,
\begin{equation}
\begin{split}
    \mathcal{L}_{\nabla} = \frac{1}{2 \tilde{H} \tilde{W}} \sum_{i=1}^{\tilde{H}} \sum_{j=1}^{\tilde{W}} \bigg[\nabla (b \odot \gls{view2d}_{T_m}) -& \\ \bigg( \nabla ( \gls{view2d}_{T_d} \odot \gls{view2d}_{T_m}) + \nabla ( \gls{view2d}_T \odot \gls{view2d}_{T_m} ) \bigg) \bigg] \\
\end{split},
\end{equation}
where $\nabla$ denotes the Laplacian gradient operator, and $\tilde{H}$ and $\tilde{W}$ are the height and width of the rendered view.
$\mathcal{L}_{\nabla}$ encourages similarity in the gradients between the blended and the dilated view combined with the gradients of the original texture, promoting boundary consistency in the blending region.
In order to further stabilize the style transformation of the blended region and encourage spatial smoothness, we use the total variation loss $\mathcal{L}_{\mathrm{TV}}$ introduced by~\cite{mahendran2015understanding}.
}
The user can control the visual appearance of the blended skin conditions by modifying the weights for each related loss term, i.e., $\lambda_c$, $\lambda_s$, $\lambda_{\nabla}$, and $\lambda_{TV}$.
Since only the areas within the masks are changed during the optimization, we use \gls{viewMask2d} to compute a padded bounding box around the skin condition and compute the loss only over this region.

Finally, we perform an iterative optimization to minimize \eqnref{eq:blend}.
At each step of the optimization, we add a small random value to \gls{weightDistance} in \eqnref{eq:camera_position}, perturbing the camera viewpoint, which helps ensure that the blended image is robust to different camera viewpoints.
We highlight that the loss function $\mathcal{L}(\cdot)$ measures the quality of the blending using the 2D views, while we optimize the underlying texture image \gls{blendedTextureImage}.
Using a differentiable renderer (\eqnref{eq:renderer}), we can calculate the gradients with respect to the pixel values of the underlying texture image, which enables the backpropagation of the loss gradients to optimize for the pixel value adjustments.

\begin{figure}[!h]
    \centering
    \includegraphics[width=\linewidth]{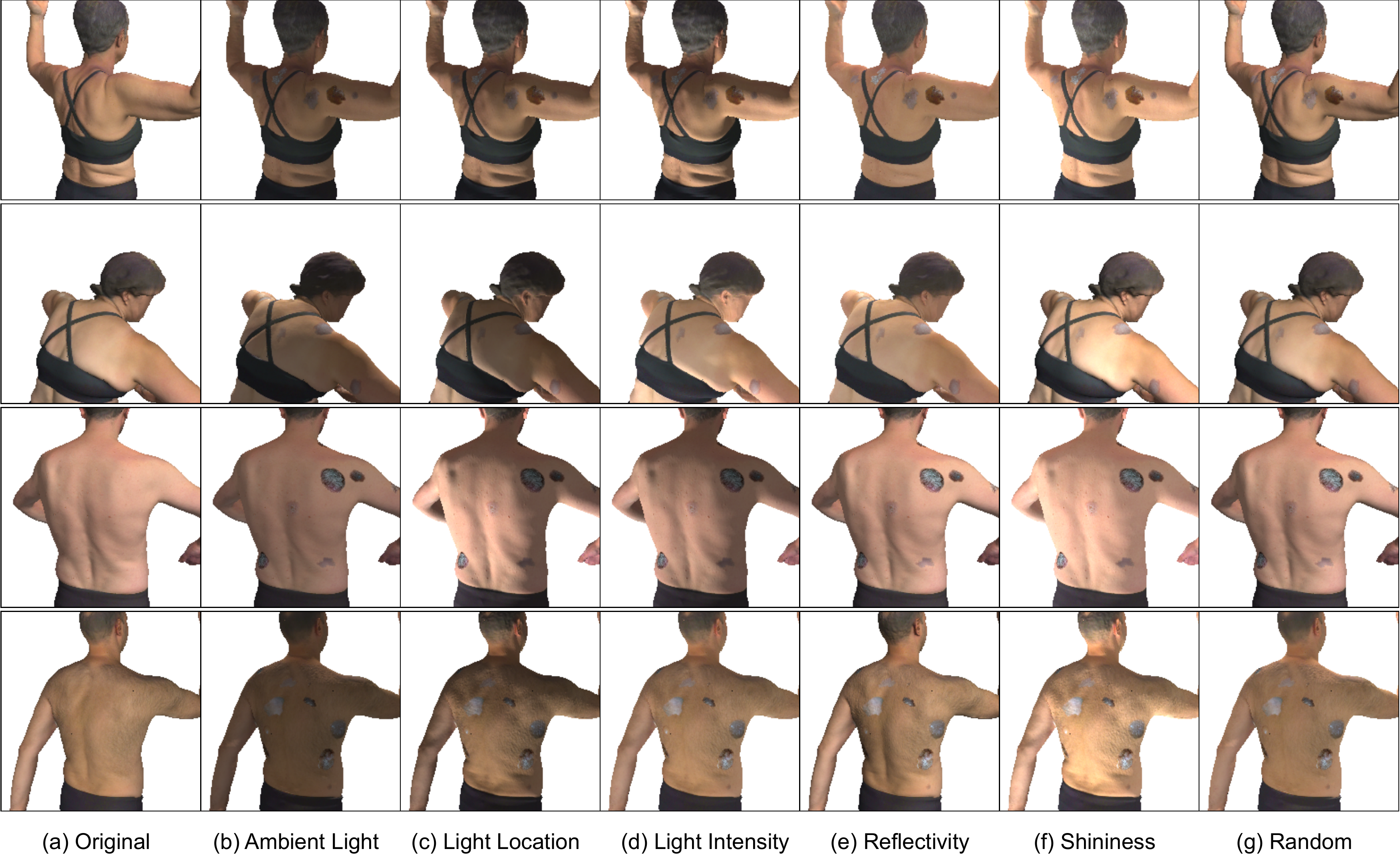}
    \caption{\ash{Rendered images from the same camera viewpoint (4 examples; one per row), showcasing blended lesions (b-g) on the original texture map (a).
    Column (a) shows the rendered views of the original texture map with a combination of Ambient, Diffuse and Specular color values for Point Lights.
    The images in columns b-d are rendered under different light source positions (b and c) and  intensities (c and d), while keeping the material properties constant.
    The images in columns e-g are rendered by changing the material's reflectivity (e), shininess (f), and a combination of both (g), while keeping the lighting parameters same as d.}}
    \label{fig:outputs_fixedpos}
\end{figure}
\subsection{Creating the 2D Image Dataset}
\label{sec:blend_dataset}

Creating the dataset of 2D rendered images and corresponding dense annotations involves two steps.
First, we determine an appropriate location for blending (\secref{sec:blend_locations}) and blend the selected skin conditions (\secref{sec:blending}) onto the texture image \gls{textureImage} of the 3D mesh \gls{mesh}. We sample a 2D image \gls{image} with skin condition from a set of real dermatological images (\secref{sec:3dbodytex}) along with an annotated mask \gls{segMask}. We apply the Shades of Gray algorithm~\cite{finlayson2004shades} to improve the color constancy within \gls{image}.
We repeat the process described in \secref{sec:blend_locations} and \secref{sec:blending} to blend $k$ skin conditions at different locations.
The output of this first step produces a blended texture image $\gls{blendedTextureImage} \in \mathbb{R}^{\gls{textureWidth} \times \gls{textureHeight} \times 3}$ and a corresponding texture mask $\gls{textureMask} \in \mathbb{Z}^{\gls{textureWidth} \times \gls{textureHeight}}$ indicating the locations of the skin conditions, where $\gls{textureWidth}$ and $\gls{textureHeight}$ are the width and the height of the original texture image \gls{textureImage} respectively. Therefore, now we have a 3D mesh with a blended lesion.

\begin{figure}[htb]
    \centering
    \includegraphics[width=\linewidth]{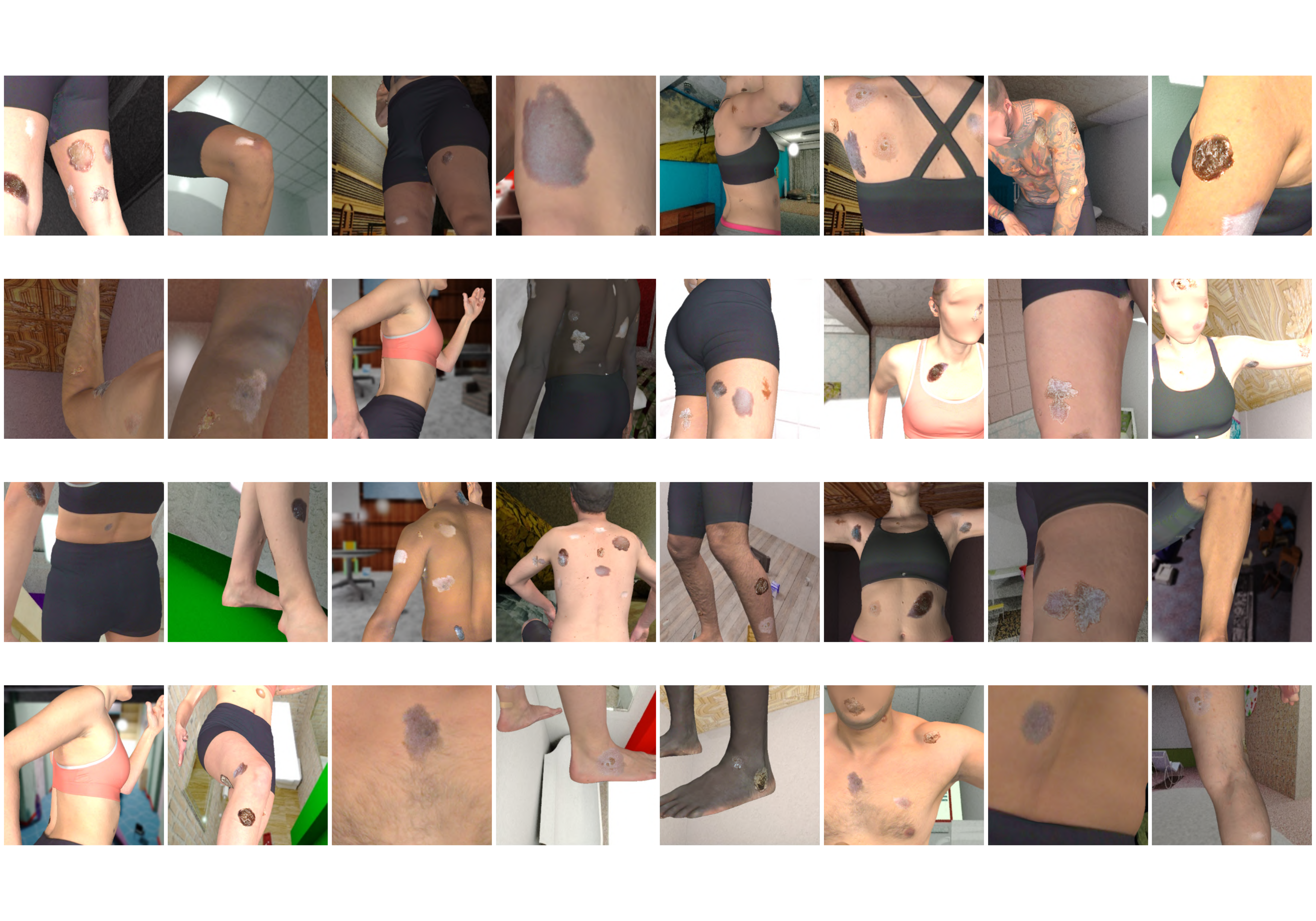}
    \caption{Generated synthetic images of multiple subjects across a range of skin tones in various skin conditions, backgrounds, lighting, and viewpoints.}
    \label{fig:example_outputs}
\end{figure}
Second, we use the blended texture image \gls{blendedTextureImage} and texture mask \gls{textureMask} to create a dataset of rendered 2D views and corresponding target labels.
To choose the camera position and direction, we use the same procedure in \eqnref{eq:camera_position} where we randomly sample a surface coordinate $\gls{cameraParameters}_{s}$ on the mesh. %
We use \eqnref{eq:renderer} with the blended texture image \gls{blendedTextureImage} to render a 2D RGB view $\gls{viewBlend2d} \in \mathbb{R}^{\gls{viewWidth} \times \gls{viewHeight} \times 3}$ and randomly sample from a range of diffuse, ambient, and specular lighting parameters and lighting positions to introduce variations in the rendered 2D views.
Moreover, for more realistic views and improved illumination, we enforce that the camera is placed outside of the mesh and that the light source reaches the camera without being blocked by the mesh.  %
To create the final image, we combine the foreground with a background image of 2D indoor scene.

\begin{figure*}[htb]
    \centering
    \includegraphics[width=0.9\textwidth]{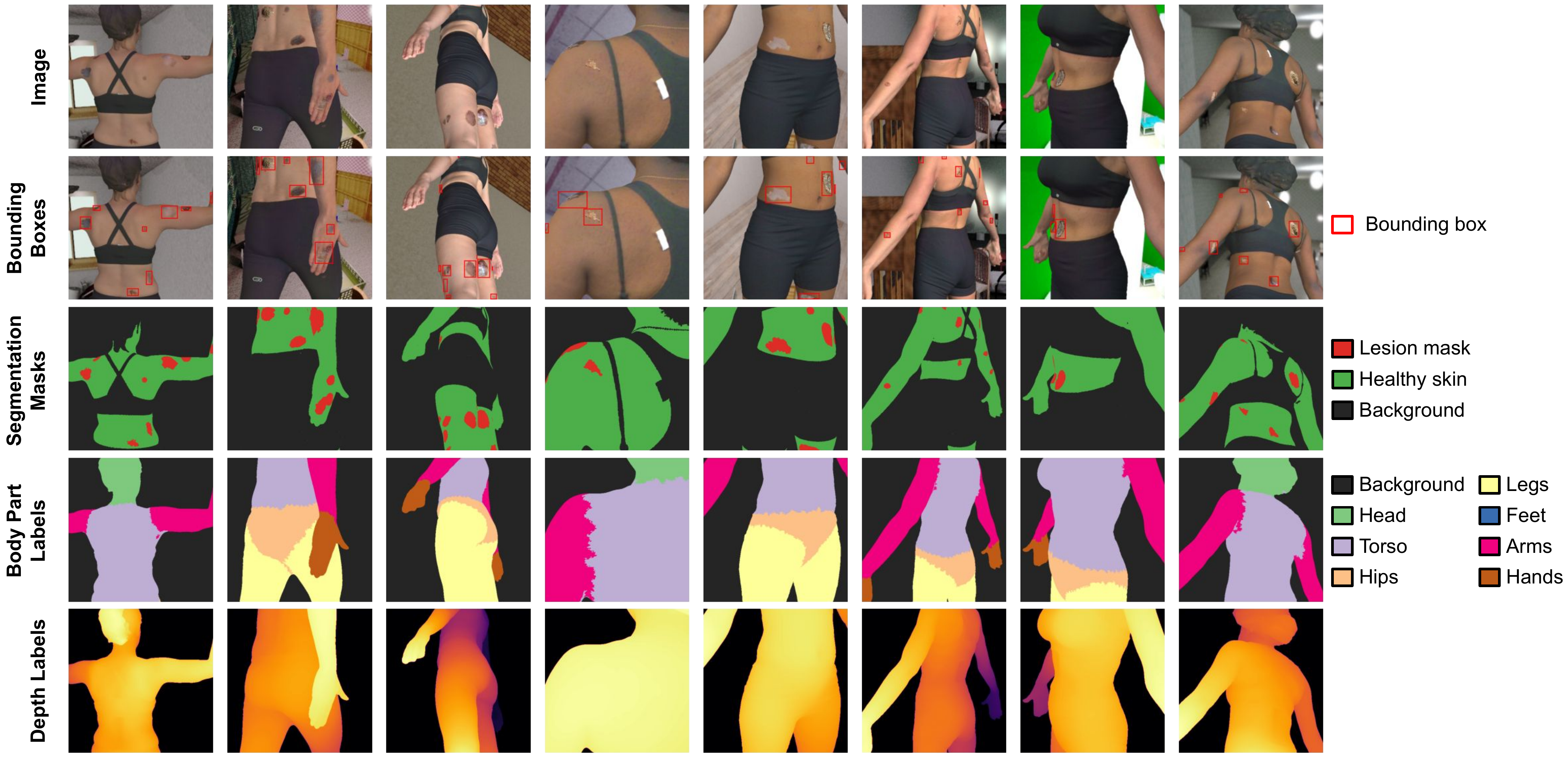}
    \caption{A few examples of data synthesized using \method{}. The rows from top to bottom show respectively: the rendered images with blended skin conditions, bounding boxes around the lesions, GT semantic segmentation masks, grouped anatomical labels, and the monocular depth maps produced by the renderer.}
    \label{fig:AnnotationOverview}
\end{figure*}

Next, we describe each of our different target variables. When rendering binary masks in \eqnref{eq:renderer}, we only use ambient lighting, as ambient light provides a uniform level of illumination to all parts of the object and preserves the underlying pixel values in the texture masks.
The skin condition mask \gls{viewMask2d} is computed by rendering with the texture mask \gls{textureMask}. The skin mask \gls{viewSkinMask2d} is computed by excluding both the skin condition regions \gls{viewMask2d} and the regions of the body labeled as non-skin (computed via the rendered view using the non-skin texture mask \gls{nonSkinTextureImage} as described in \secref{sec:3dbodytex}). The non-skin mask \gls{viewNonSkinMask2d} is computed from regions containing neither skin \gls{viewMask2d} nor skin conditions \gls{viewSkinMask2d} (Figure~\ref{fig:AnnotationOverview}, third row).
Additionally, we obtained bounding boxes around skin condition regions by computing the minimal enclosing box around each skin condition mask (Figure~\ref{fig:AnnotationOverview}, second row from the top).

In addition to the skin lesion segmentation masks, we create other ground truth annotations, such as body part labels and depth maps. For the body part labels, we include dense anatomical labels for 16 regions of the body (e.g., head, torso) where we use the mesh's faces \gls{faces2d} (from \eqnref{eq:renderer}) visible in the rendered view \gls{viewBlend2d} to determine an anatomy label for each pixel in the rendered view (\secref{sec:anatomy} describes how a mesh is assigned anatomical labels) and assign a background label to pixels outside the mesh (Figure~\ref{fig:AnnotationOverview}, fourth row from the top).
For the depth maps, the depth image \gls{viewDepth2d} is obtained from the fragments computed by the renderer (Figure~\ref{fig:AnnotationOverview}, bottom row).

Finally, we generate our dataset by rendering a set of 2D images and the corresponding annotations for each mesh, by sampling $n$ times under different camera, lighting, and material parameters, and background scenes.
\ash{We show some example images from the generated 2D dataset in \figref{fig:outputs_fixedpos} and }
\ash{\figref{fig:example_outputs}.}

\begin{figure}[hbt]
    \centering
    \includegraphics[width=\linewidth]{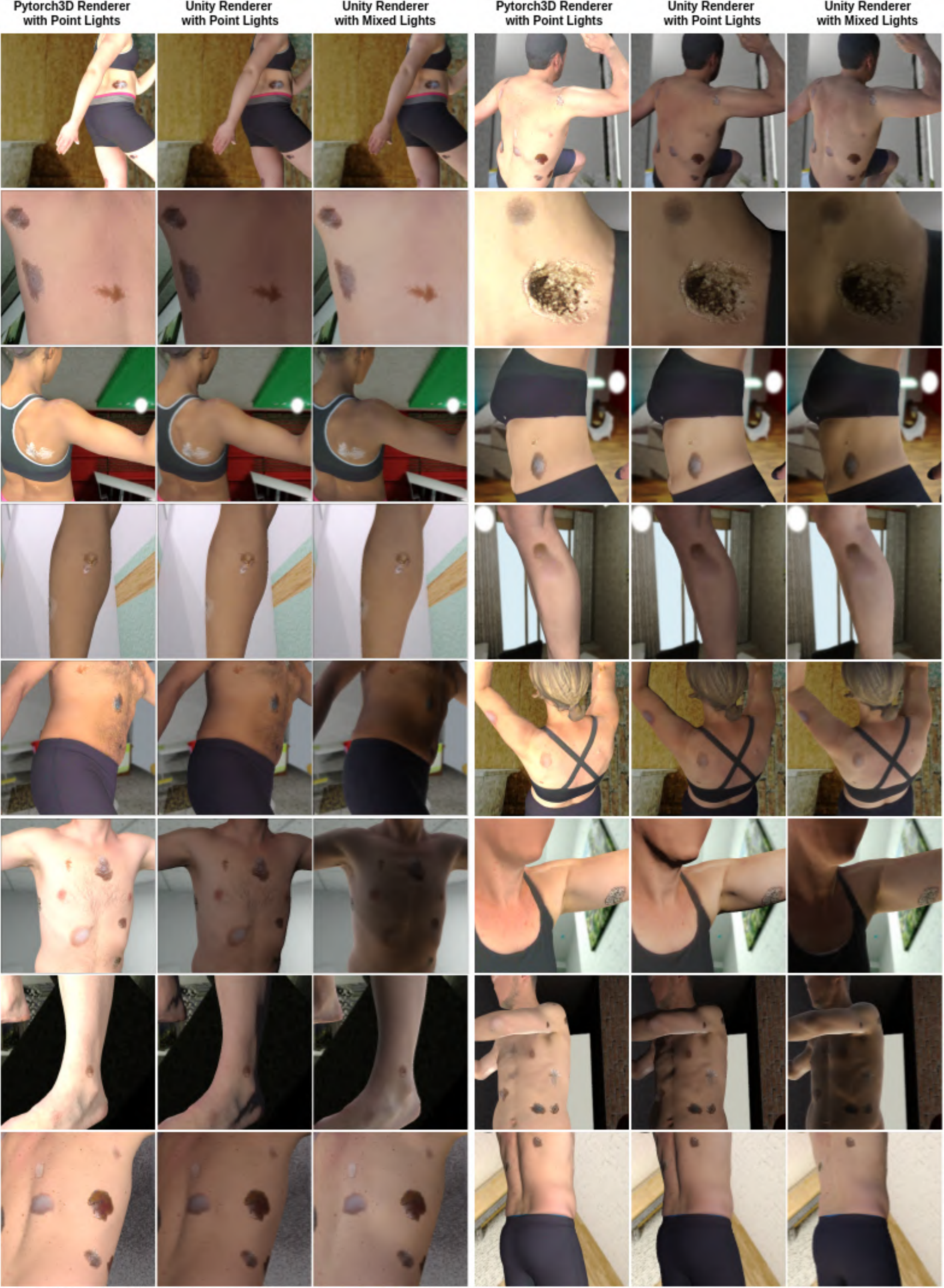}
    \caption{Some samples of the 2D images generated using PyTorch3D~\cite{Ravi2020}, and Unity3D renderer.
    We show the rendered images from Unity3D in two lighting conditions, namely Point Lights and a mix of Point and Direction Lights.
For the experiments reported in the paper, we use PyTorch3D with Point Lights since it mimics the natural behavior of light in the real world.}
    \label{fig:renderComp}
\end{figure}

The modular design of our \method{} pipeline not only allows us to easily modify the aforementioned settings, but also allows us to achieve photo-realistic rendering by replacing the differential renderer $R(\cdot)$ with any physically based rendering (PBR) method such as Unity3D\footnote{https://docs.unity3d.com/ScriptReference/Renderer.html}
However, for all the experiments reported in the paper, we use PyTorch3D~\cite{Ravi2020} owing to its simplicity and wide adoption in the research community.
We show some qualitative samples obtained using these two kinds of renderers in \figref{fig:renderComp}.
We observe that regardless of the choice of the renderer used in creating the 2D dataset of rendered images, the downstream task of foot ulcer wound segmentation achieved almost similar quantitative performance, as shown in \figref{fig:renderSeg}.

\section{Materials: Datasets and Annotations}

\begin{figure}[htbp]
    \centering
    \includegraphics[width=\linewidth]{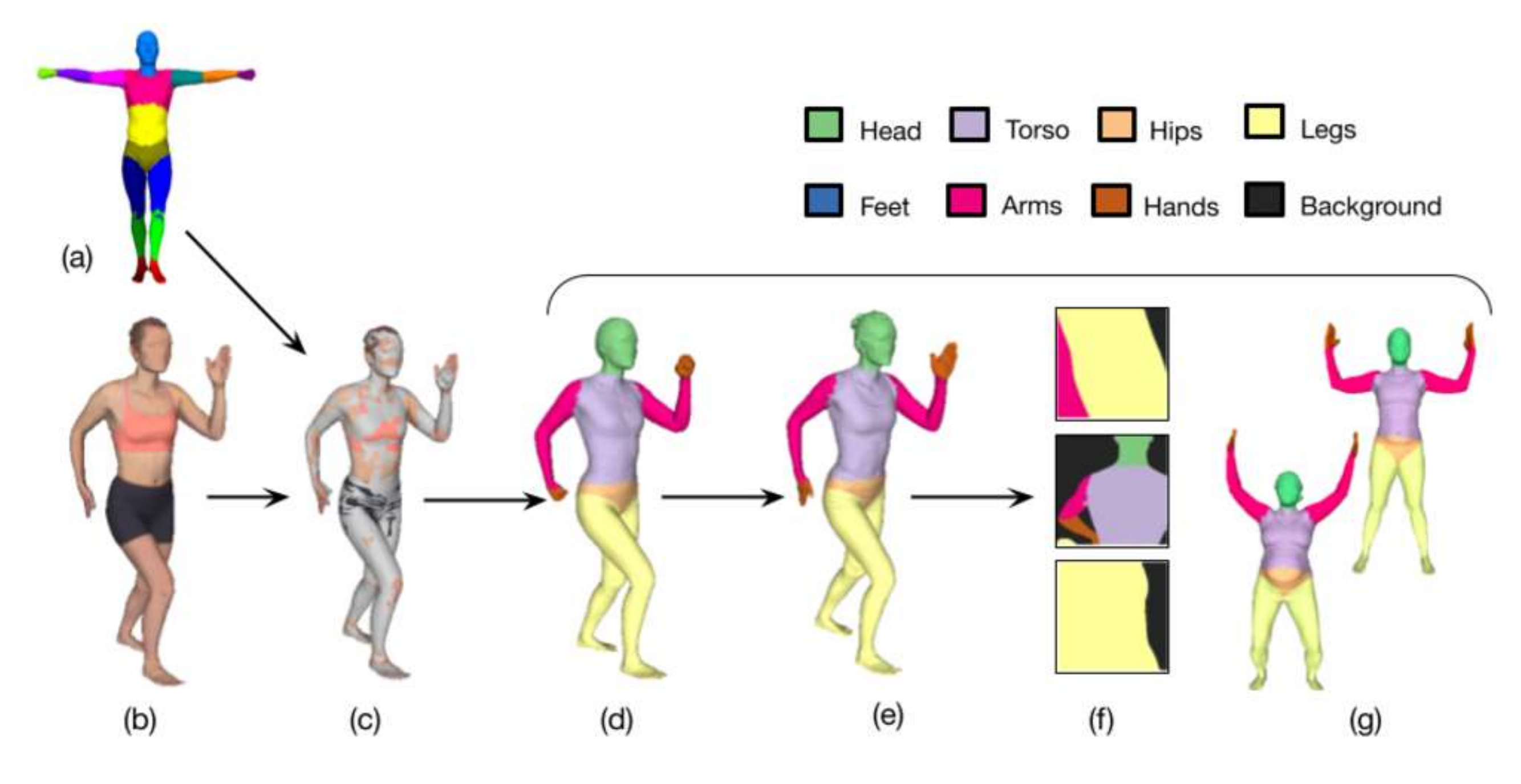}
    \caption{Proposed schema for anatomical part labeling showing: (a) 3D SCAPE template body model with $16$ anatomical part labels, (b) Input 3DBodyTex scan, (c) Template model fitted to input scan, (d) Anatomical body part labels (only 7 instead of 16, as per~\cite{parsing_tested}) assigned to fitted model, (e) Anatomical labels transferred to the input scan, (f) 2D views of annotated samples, (g) More examples. }
    \label{fig:3D_body_part_assignment}
\end{figure}

\subsection{3D Textured Human Meshes and Their Annotation}
\label{sec:3dbodytex}

We use the 3DBodyTex~\cite{Saint2018, Saint2019} dataset, which provides 400 high-resolution textured meshes from 200 unique subjects, each imaged in two different poses.
Subjects wear sports clothing, which results in a significant amount of skin regions being exposed and captured.
We manually annotate non-skin regions within the 2D texture image to create a non-skin texture mask \gls{nonSkinTextureImage}, which we define as regions containing clothing, hair, jewelry, etc.
We select a subset of 50 annotated meshes to perform blending, where meshes are chosen to
cover a wider range
of skin tones available within 3DBodyTex.
We provide further details in \ref{sup:sec:annotate_nonskin}.

\subsection{Anatomy Labels for 3DBodyTex}
\label{sec:anatomy}

We segment the 3D body mesh \gls{mesh} into different anatomical parts by fitting the SCAPE body model~\cite{scape2005} with $16$ anatomical parts annotated per-vertex onto \gls{mesh} using an automatic fitting method described by~\cite{Saint2019}.
This yields a fitted mesh with the same topology and geometry as the body model, thus inheriting the body part annotations.
As the fitted mesh has the same shape and pose as \gls{mesh}, we can transfer the body part annotations using the nearest vertex assignment between the fitted mesh and \gls{mesh}.

In \figref{fig:3D_body_part_assignment}, we show the process of labeling given an input 3D body scan and a SCAPE~\cite{anguelov2005scape} template body model.
Initially, the body model comes with 16 anatomical labels consisting of: \textit{head, upper torso, lower torso, hips, upper leg left, upper leg right, lower leg left, lower leg right, feet left, feet right, upper arm left, upper arm right, lower arm left, lower arm right, hand left, hand right}.
For simplicity and a fair comparison with~\cite{parsing_tested,chen2014detect}, only 7 anatomical labels are kept namely: \textit{head, torso, hips, legs, feet, arms, and hands} (\figref{fig:3D_body_part_assignment}-(g)).

\subsection{Segmentation of 2D Dermatological Images}
\label{sec:fit17k}

We use the Fitzpatrick17K dataset~\cite{Groh2021}, a clinical dataset composed of 2D ``in-the-wild'' clinical images and corresponding disease labels.
We manually segment a total of $75$ images into lesion, skin, and  background segmentations, where 50 images were used for blending and as a validation set during training, and 25 images were held out for evaluation.
We point the readers to \ref{sup:sec:annotate_fitz} for more details.

\subsection{Backgrounds for Synthetic Images}
\label{sec:background}
In the final step of generating synthetic images, we combine the foreground with a background image. Since real-world clinical settings are usually indoor environments, for generating 2D ``in-the-wild'' clinical images, we randomly choose the background from publicly available 2D indoor scene images~\cite{McCormac2017, ReniHouseRoomDataset}.

\section{Implementation Details}
\label{sec:implementation}

\subsection{Dataset Construction Details}
\label{sub:sec:dataset_const}

\subsubsection{Placing and Blending the Skin Condition into a Mesh}
For the rendering in \eqnref{eq:renderer}, we set the rendered width \gls{viewWidth} and height \gls{viewHeight} to $512 \times 512$, and set the lighting parameters \gls{lightingParameters} to use only point lights, placed at the same position as the camera, to avoid interference from shadows.

During the blending stage, the RGB components of ambient, specular, and diffused colors of the lighting parameters are set to $0.5$, $0.025$, and $0.5$ each, respectively.
We also fix the specular color and shininess of the material to be $0.025$ and $50$, respectively.
These values determine the intensity of luster and the color of the reflected light from the material, i.e., the texture image.
Additionally, we use perspective projection-based cameras and set the field-of-view (FOV) as $30\degree$.

To determine the surface coordinate $\gls{cameraParameters}_{s}$ in \eqnref{eq:camera_position}, we explored using the center of a sampled mesh face \gls{face} and sampling points on the surface of the face and found minimal differences.
However, this is likely dataset dependent, where meshes with coarse non-uniform faces benefit more from sampling surface points with a probability proportional to the area of the face.
We set \gls{weightDistance} to a random value in the range [0.4, 0.6], which is a dataset-dependent range we empirically set that contributes to the scale of the blended lesion.
When placing multiple skin conditions into a single texture image, we apply random horizontal and vertical flips and rotations to the lesions to introduce variability in the orientation of the lesion, where the texture mask keeps track of where the skin condition is blended to prevent skin conditions from overlapping.

To minimize \eqnref{eq:blend}, we use the Adam optimizer~\cite{Kingma2015} with a learning rate of $0.005$ and optimize for $400$ steps per location. In \eqnref{eq:blend_loss}, we set weights for each loss function as $\lambda_c = 2$, $\lambda_s = 10^6$, $\lambda_{\nabla} = 10^5$, and $\lambda_{\mathrm{TV}} = 10^{-4}$, which are set empirically to blend the lesions into the textures while preserving many of the lesions' visual characteristics.

\subsubsection{Rendering 2D Views and Creating the Dataset}
We fix all rendered blended views to a height \gls{viewHeight} and width \gls{viewWidth} of $512 \times 512$. We sample from a range of camera views and lighting parameters, where the ranges are empirically set to span across a variety of plausible ``in-the-wild" acquisition scenarios. We sample \gls{weightDistance} between [0.1, 1.3] to give a range of closeup and full body field-of-views.
To introduce a variety of lighting conditions while creating the 2D dataset, we sample the RGB components of ambient and diffused colors in lights from a range of [0.2, 0.99], whereas the specular color is sampled between [0, 0.1].
Furthermore, the specular color and shininess of the material is sampled between a range of [0, 0.05] and [30, 60], respectively.

\begin{figure*}[htb]
     \centering
     \begin{subfigure}[b]{0.40\textwidth}
         \centering
         \includegraphics[width=0.90\textwidth]{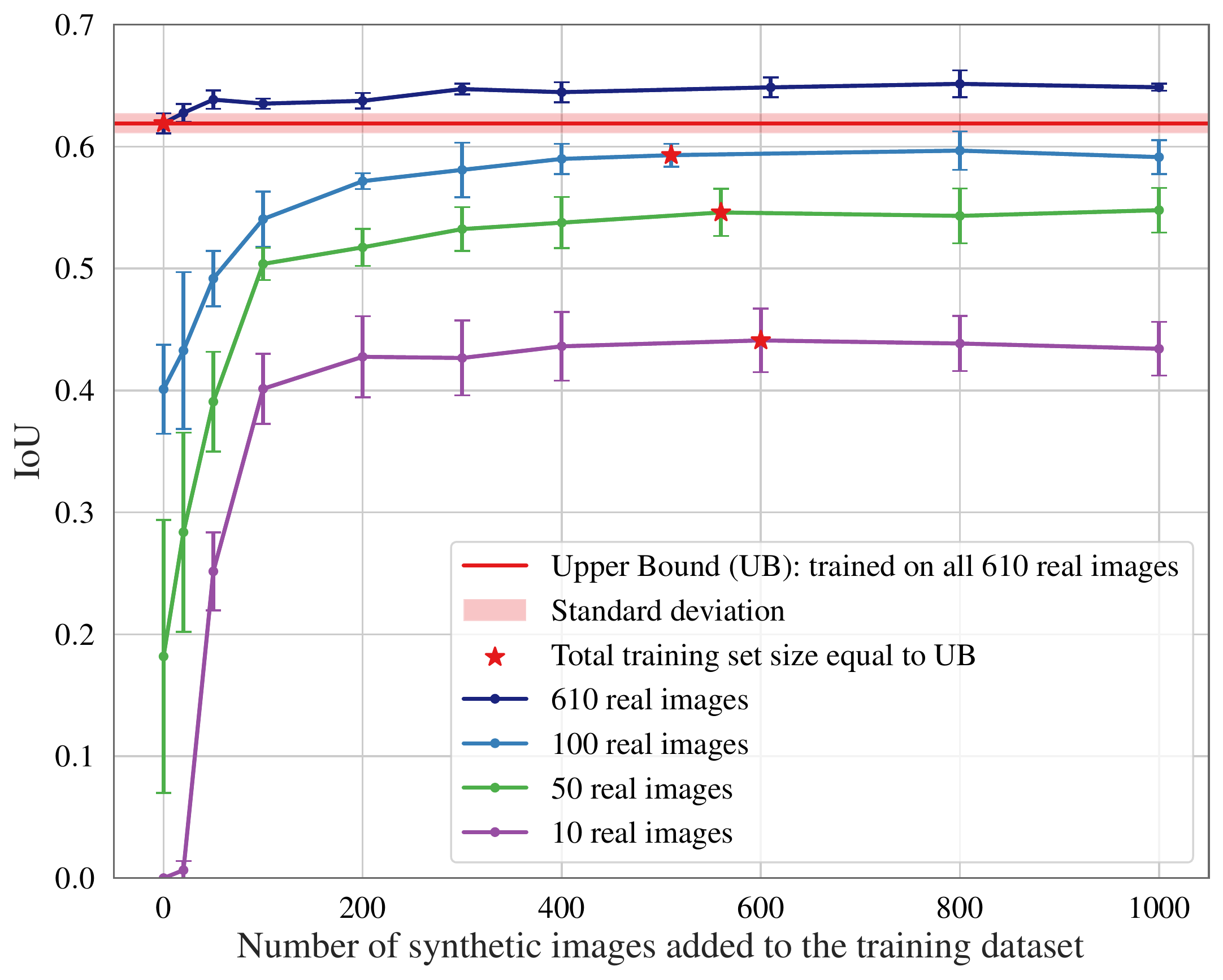}
         \caption{}
     \end{subfigure}
    \hspace{2.5em}
     \begin{subfigure}[b]{0.40\textwidth}
         \centering
         \includegraphics[width=0.90\textwidth]{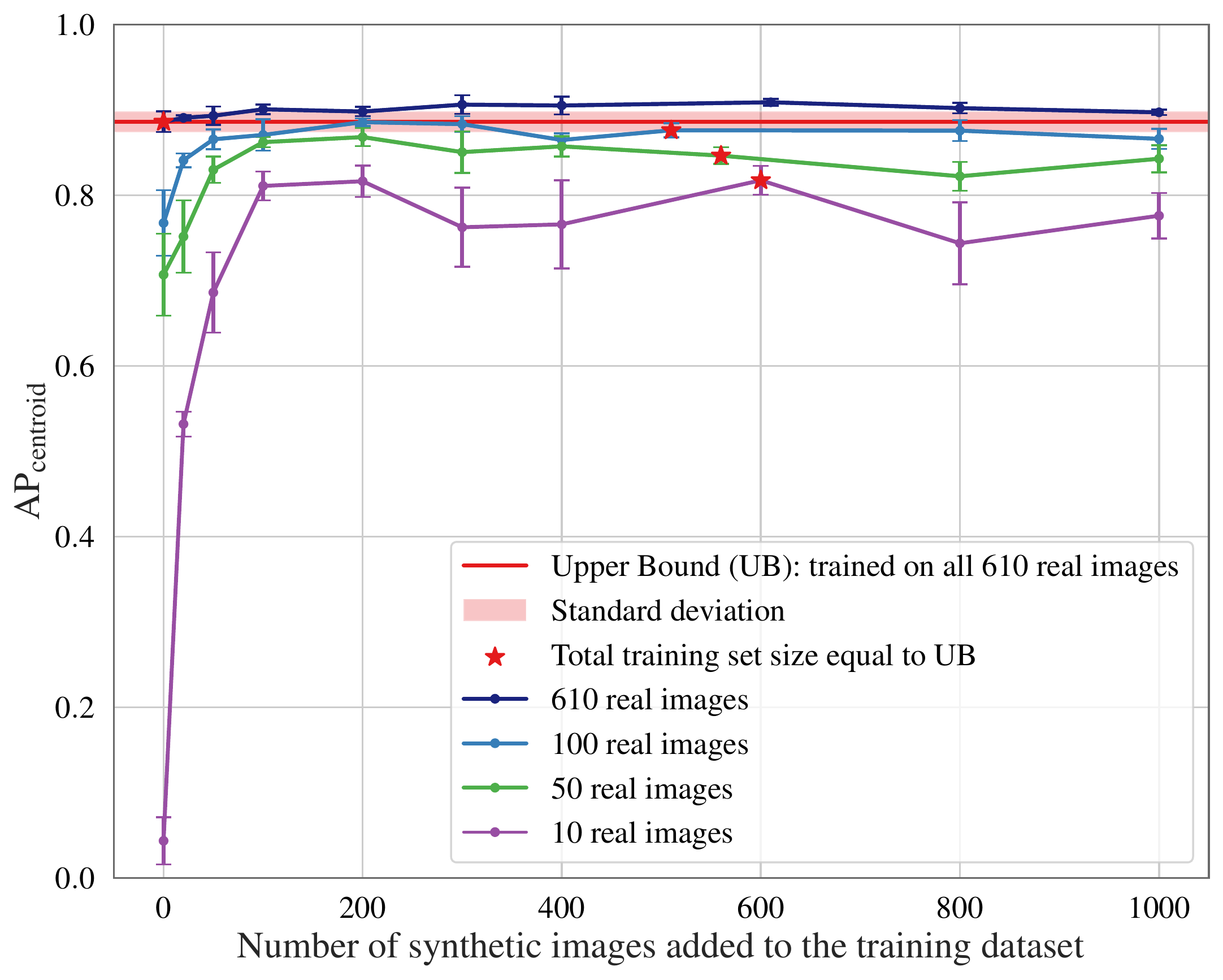}
         \caption{}
     \end{subfigure}
        \caption{Wound bounding box detection performance across five folds (mean and standard deviation) on FUSeg dataset, where the number of synthetic images added to a fixed number of real images in the training set gradually increases. Bounding box detection performance is measured by (a) IoU and (b) $\mathrm{AP}_{\mathrm{centroid}}$ (note that the vertical scales of the two plots are different). \ari{The plotted results extend up to the point of convergence.} The horizontal red line indicates the results for the model that is trained on 610 real images,
        \ash{which shows the bounded performance using all the real images.}}
        \label{fig:synth_augment}
\end{figure*}

\subsection{Experimental Details}
\subsubsection{Wound Bounding Box Detection and Semantic Segmentation in Clinical Images}
\label{sub:sub:sec:wound_exp}
We use the FUSeg dataset from the \emph{The Foot Ulcer Segmentation Challenge}~\cite{Wang2020}, which contains 2D clinical dermatological images of ulcers on the foot and the corresponding wound masks.
The FUSeg dataset contains the standard training, validation, and testing partitions of 810, 200, and 200 images, respectively.
As the ground truth annotations for the official test set are not publicly released, we use the official validation set for our evaluation and split the official training set into 610 images for training and 200 images for internal validation.

For the wound detection task, we convert the masks of the wounds to bounding boxes by labeling the connected regions of the masks and computing the minimal enclosing bounding box, and train
a Faster R-CNN~\cite{Ren2016} model for bounding box detection.
We use a mini-batch size of 8 images and train the model for a maximum of 50 epochs using SGD~\cite{Robbins1951,kiefer1952stochastic,Bottou2018} with a learning rate of $0.001$.
We choose the model weights with the maximum intersection over union (IoU) score over the internal validation set of real images.

For the wound segmentation task, we train a DeepLabV3~\cite{Chen2017} network with a ResNet-50~\cite{he2016deep} backbone as our model. We use a mini-batch size of 8 images and minimize the binary cross entropy loss for a maximum of 250 epochs using the Adam optimizer with a learning rate of $0.00005$ and a weight decay of $0.00005$.
We choose the model weights with the maximum Dice score over the internal validation set of real images.

\subsubsection{Lesions, Skin, and Background Segmentation Using In-the-wild Clinical Images}
For this experiment, we train a DeepLabV3 ResNet50~\cite{Chen2017} CNN model for a maximum of 3,430 steps, using a batch size of 8.
We train on 13,720 synthetic images while validating using 50 real 2D dermatological images, and use a modified fuzzy Jaccard index~\cite{Crum2006} as the loss function.
See \ref{sup:sec:lesion_skin_bg} for more details on the loss and the training procedure.
We choose a model that produces the lowest Jaccard loss on our validation set consisting of 50 real 2D dermatological images
that we manually segmented into regions for skin conditions, healthy skin, and non-skin.
The CNN training process on this task took around one hour on an NVIDIA GeForce RTX 2070 Super 8 GB GPU, while generating the synthetic image dataset after blending took approximately three hours.
Creating the dataset with 50 meshes, each with 50 skin conditions blended onto them, took around 30 hours on a Quadro-RTX-4000 8GB GPU.
 It is worth noting that the majority of the computing time was spent on blending the lesions to generate the dataset.

\subsubsection{Lesion and Skin Segmentation on Dermoscopy, Clinical, and Non-Medical Images}
In \secref{sec:results:segment:other}, we describe experiments using the dermoscopy dataset PH2~\cite{Mendonca2013,Ferreira2012}; the clinical dataset DermoFit~\cite{Ballerini2013}; and the non-medical skin dataset Pratheepan~\cite{Yogarajah2010,Tan2012}.
We pre-process these images by applying the Shades of Gray~\cite{finlayson2004shades} color constancy algorithm followed by image normalization (i.e., subtract image pixels by the pretrained dataset channel mean and divide by the channel standard deviation). Images are resized to maintain their aspect ratio such that the smallest spatial dimension is equal to the spatial dimension of the resized training images (320 pixels), expect for the dermoscopy dataset PH2, which is resized based on the longest spatial dimension (PH2 contains dermoscopy images, which show dermoscopic structures not visible to the naked eye and these types of images were not part of the training data). We apply Gaussian smoothing to the PH2 and DermoFit images as these images show a close-up view of the lesion with details that may not be visible in our training data. %

\begin{figure*}[hbt]
     \centering
     \begin{subfigure}[b]{0.49\textwidth}
         \centering
         \includegraphics[width=0.90\textwidth]{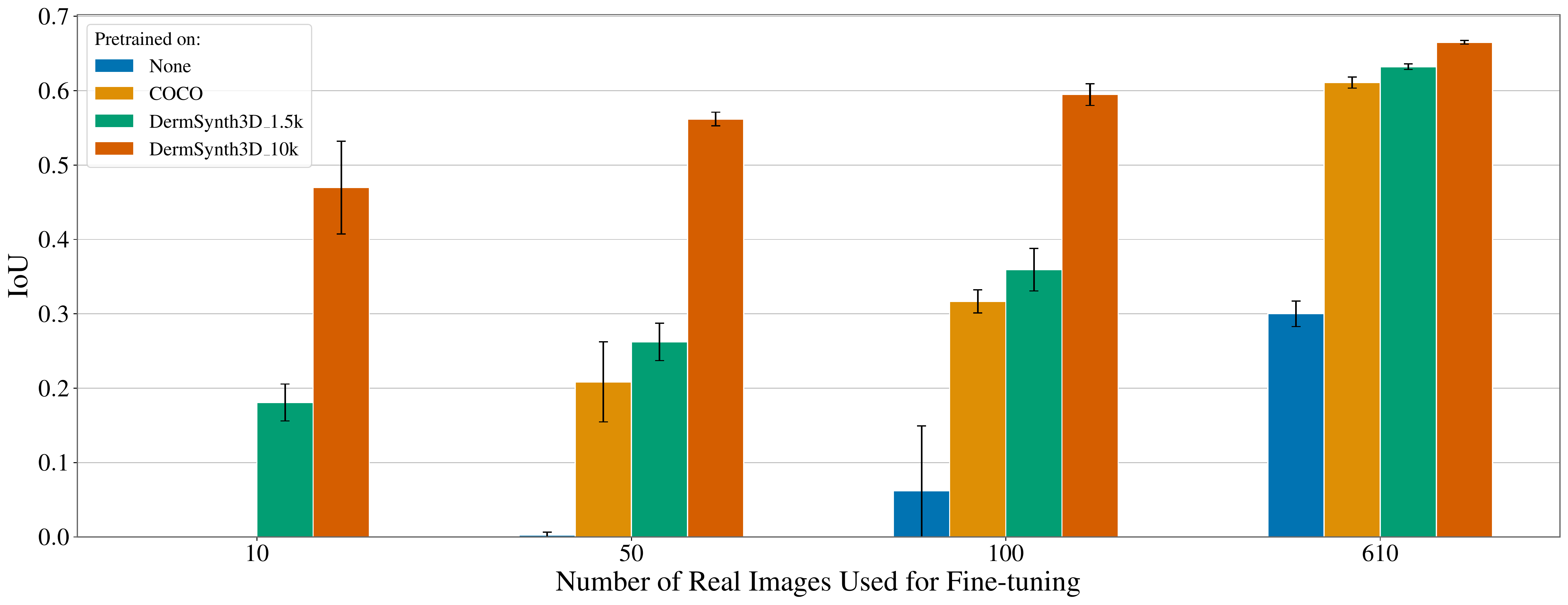}\\
         \vspace{1.5em}
         \includegraphics[width=0.90\textwidth]{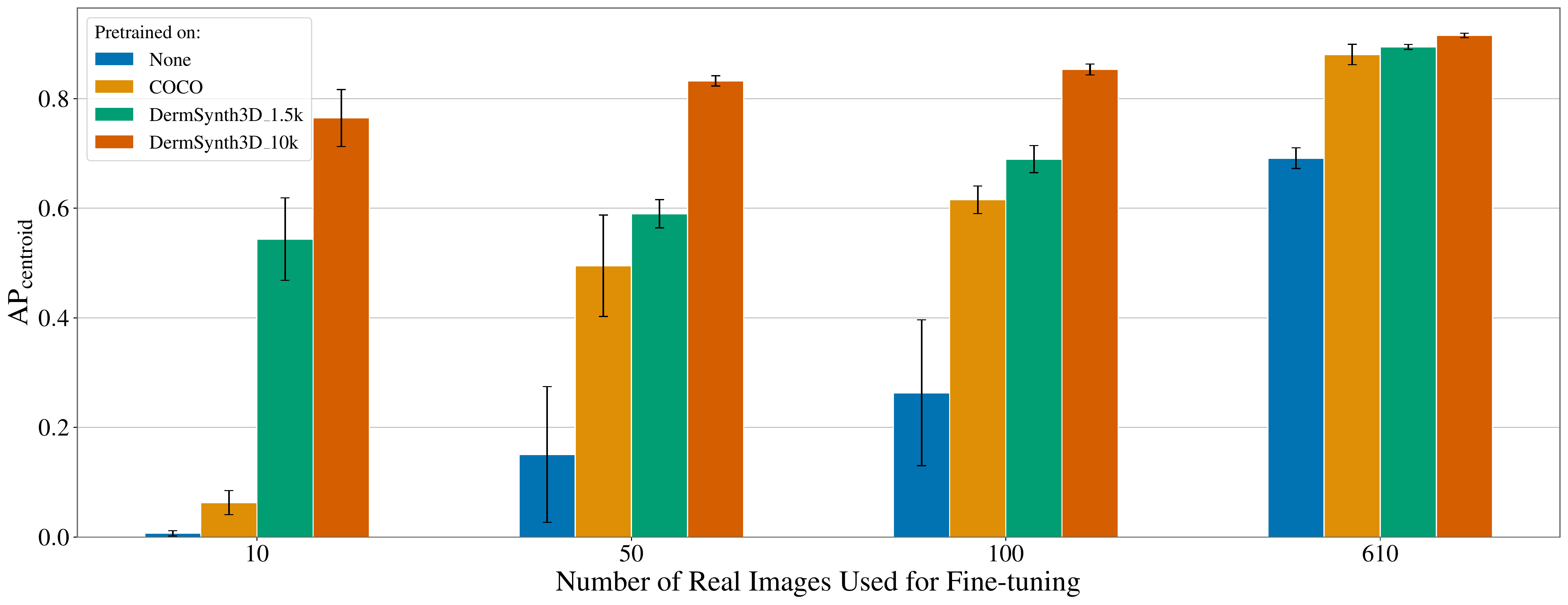}
         \caption{Foot Ulcer Detection}
     \end{subfigure}
     \begin{subfigure}[b]{0.49\textwidth}
         \centering
         \includegraphics[width=0.90\textwidth]{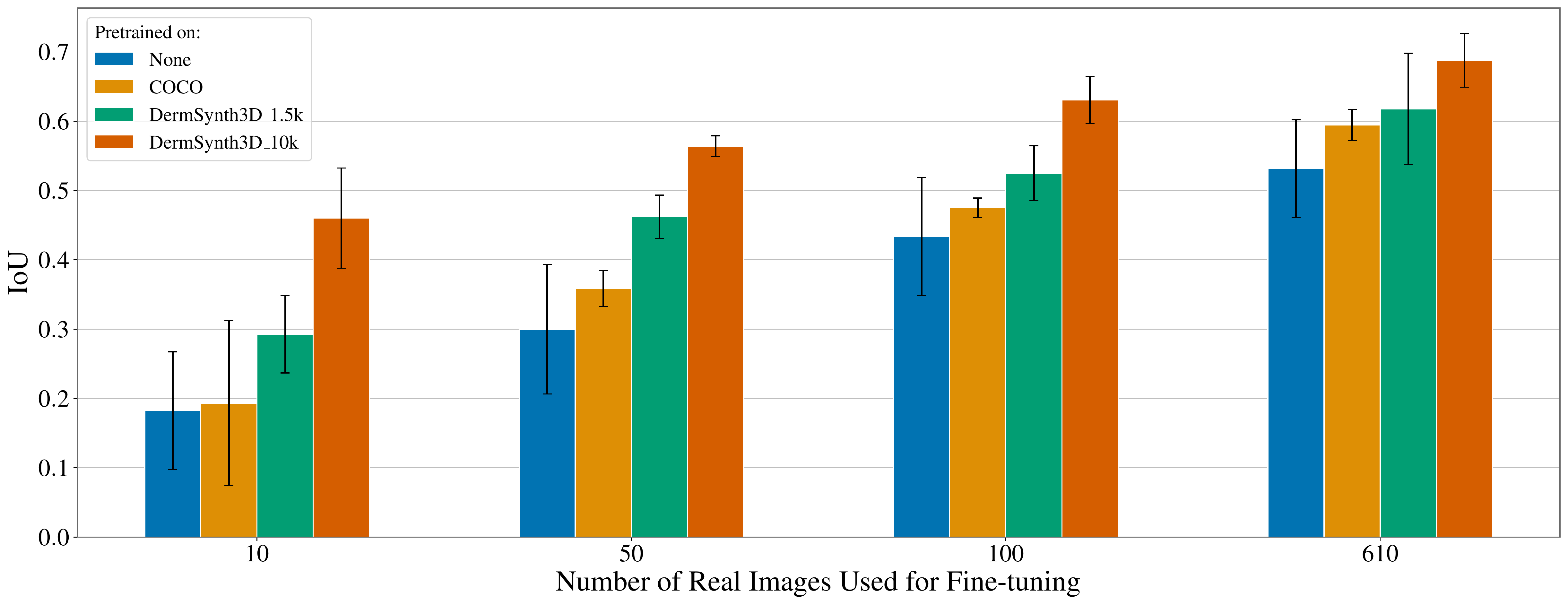}\\
         \vspace{1.5em}
         \includegraphics[width=0.90\textwidth]{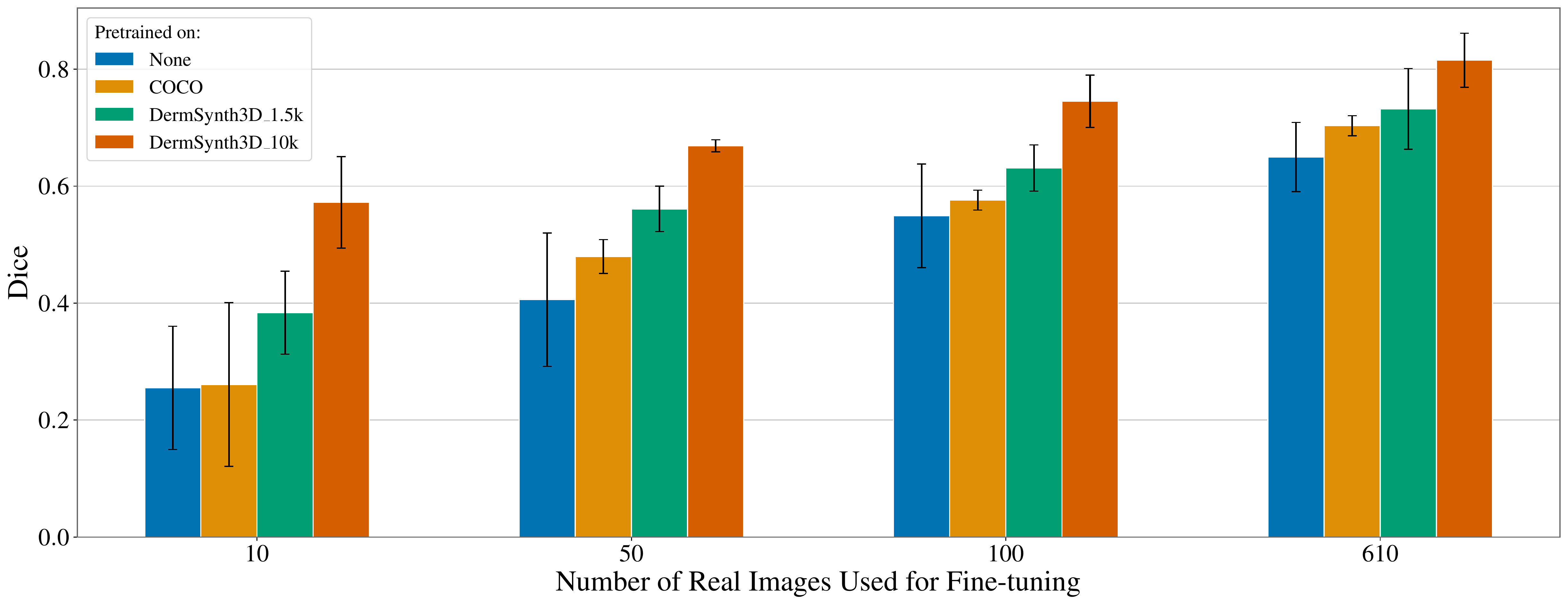}

         \caption{Foot Ulcer Segmentation}
     \end{subfigure}
        \caption{\ash{
        Comparative analysis on the effect of pre-training on COCO dataset and synthetic data generated by \method{}, in the context of two downstream tasks, namely lesion detection and segmentation. We compare the performance of the models that are: $(i)$ trained from scratch \ie, ``None", $(ii)$ pretrained on COCO, $(iii - iv)$ pretrained on $1,500$ and $10,000$ synthetic data produced by \method{}. We observe a substainal increase in performance during fine-tuning on real data when using models pretrained on \method{}, as opposed to COCO. We also notice the benefits of pretraining on large-scale data. These results highlight the practical advantage of using \method{} in enhancing the model's generalization to real data.
        }
        }
        \label{fig:pretrain}
\end{figure*}
\section{Experiments and Results}
\label{sec:experiments}

We train deep learning models with pre-trained weights for bounding box detection and semantic segmentation on our synthetic data using common image augmentation and normalization techniques (e.g., rotation, color shifts), and evaluate on real 2D images including dermatological images with skin conditions.
We perform these experiments in order to evaluate how well a model trained on our generated synthetic data can generalize to unseen real data.
We emphasize that our goal in these experiments is not to compete with state-of-the-art performance over these datasets, but rather to show the utility of the generated dataset by assessing the model's ability to generalize to real 2D images when trained on this dataset.
Ideally, we would evaluate our approach over an existing ``in-the-wild'' clinical dermatological dataset with skin conditions, skin, and background segmentation labels.
However, to the best of our knowledge, there exists no such dataset, as most skin image datasets contain labels for binary segmentation tasks (e.g., skin vs background or lesion vs background).
Thus, we evaluate using data we manually annotated over an ``in-the-wild'' clinical dataset containing skin conditions, skin, and background masks, as well as over different binary segmentation tasks on prior datasets.

\subsection{Wound Bounding Box Detection and Semantic Segmentation in Clinical Images}
\label{sec:woundsseg}
Detecting wounds in clinical images is an important step to track and extract morphological features from the wounds, which is crucial for diagnosis and treatment.
Bounding boxes can be used to localize the wounds in clinical images and minimize unnecessary information within the scene to improve downstream tasks~\cite{Wang2020}.

For evaluating the bounding box detection performance, we use two metrics: the intersection over union (IoU) score, which measures the exact match between a detected and ground truth bounding box, and the average precision of overlapping centroids ($\mathrm{AP}_{\mathrm{centroid}}$)~\cite{zhao2021skin3d}, which determines the bounding box localization performance, rather than its precise boundaries and is more suitable for medical applications.

To assess the performance improvement from using synthetic images in the training process, we gradually increase the number of synthetic images added to the training sets of limited real images.
We can see in Figure~\ref{fig:synth_augment} that \ari{augmenting the entire real training dataset with synthetic images significantly improves the wound detection performance. This observation highlights the capacity of synthetic images to introduce meaningful information (beyond what is in the real images) during training.}
\ari{ Figure~\ref{fig:synth_augment} demonstrates that} the addition of synthetic images consistently improves the detection performance and reduces the standard deviation error in the results, thus leading to more robust and reliable performance.
\ash{
We note that the performance of the model converges after the addition of 400 synthetic training images and increasing them beyond 1000 did not significantly increase the performance.
However, this maybe partly application-dependent.
}

Moreover, using only less than $\frac{1}{6}$\textsuperscript{th} of the available real images (100 annotated real images) alongside synthetic ones, we can achieve comparable detection results to the upper bound, which is less than a $2\%$ drop in performance.
Note that for generating synthetic training images using \method{}, only 50 lesion annotations were used, which is $8.2\%$ of the cost of dense annotations compared with the real dataset of wounds.
Another notable observation in Figure~\ref{fig:synth_augment} is that by adding 100 synthetic images to a very small dataset of 10 real images, we can achieve a similar performance as a dataset of 100 real images.
This demonstrates the usefulness of this approach in situations where real data is extremely limited.

To further explore the usefulness of our synthetic images in scenarios where there is no real training data available, we conduct additional experiments.
We create a synthetic dataset of 610 images, which is the same size as the ``real" wound image training set of the FUSeg dataset.
We then evaluate the performance of a model in bounding box detection and segmentation when it is trained on this \emph{synthetic-only} dataset and tested on the real wound image testset.

The quantitative results are reported in Table~\ref{table:fuseg-det-and-seg} alongside the model's performance when trained on the FUSeg training set of real wound images, under the same training settings.

Our experiments show that for wound detection, when only synthetic \method{} data is available, an average precision of $80\%$ in wound localization can still be achieved.
Additionally, for the segmentation performance, a model trained on only synthetic images still achieves a Dice score of $0.49$, which is more than $60\%$ of the performance on real data ($0.81$ Dice), despite the differences in semantic content (skin conditions selected from Fitzpatrick17K dataset versus foot ulcers) and source domains (synthetic versus real).
This demonstrates that even in the absence of real images, training on synthetic \method{} data can provide more than $60\%$ of the expected performance when trained on real clinical images, despite the significant domain gaps.

\subsubsection{\ash{Pretraining with Synthetic Data}}
\ari{
Since the introduction of AlexNet~\cite{krizhevsky2012imagenet}, utilizing models pretrained on large amounts of data and finetuning them for downstream tasks has been a popular practice in the computer vision community~\cite{sharif2014cnn, kornblith2019better}.
However, existing pretrained models are trained on natural images, which have a considerable domain gap with medical images.
One of the main reasons that models pretrained on medical images are not available, is the problem of annotation burden and the cost of creating large-scale datasets that can be used for pretraining models.
However, our proposed data synthesis framework, \method{}, has the potential to create large-scale data with a relatively much lower cost.
Therefore, we conduct additional experiments using synthetic data generated by \method{} to pretrain a model before fine-tuning it on real data.}
\ash{
As a baseline experiment, we follow the experimental settings described in~\secref{sub:sub:sec:wound_exp} of using a model pretrained on COCO dataset~\cite{lin2014microsoft} and fine-tune it on real images to segment and detect wounds, respectively.
For our second experiment, we initialize a model with random weights (i.e., without COCO-pretrained weights), train the model until convergence on (i) 1,500, and (ii) 10,000 synthetic images generated by \method{} to segment and detect lesions, and then fine-tune the model on real images.
We also consider a na\"ive approach where a model is initialized with random weights and trained on only the real images.
We evaluate performance on the FUSeg dataset of real wound images, and report the performance when varying the number of real training images (10, 50, 100, and 610 real images).
Our results (\figref{fig:pretrain}) indicate that a model pretrained on \method{}'s synthetic data outperforms a model pretrained on COCO for segmenting and detecting real wound images.
These results suggest that generated synthetic data may have a role in pretraining models, which may be especially beneficial when low numbers of real images are available.
}

\begin{figure*}[!hbt]
    \centering
    \includegraphics[width=\linewidth]{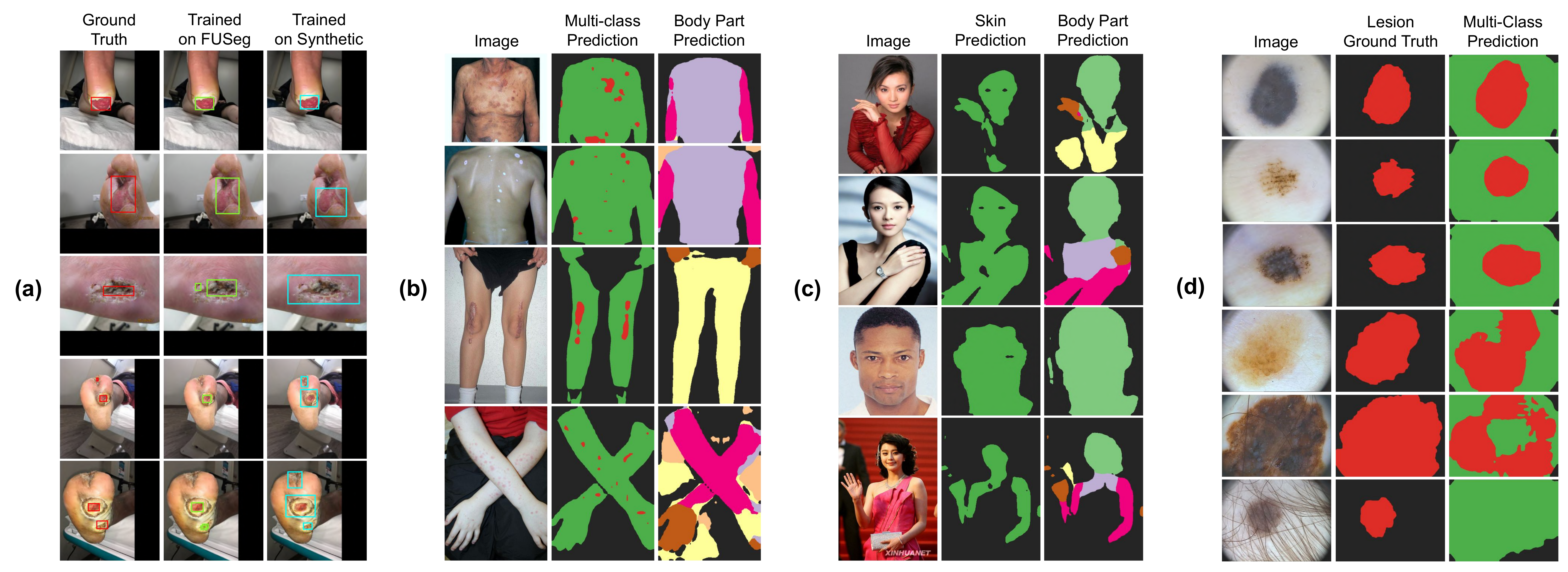}
    \caption{Qualitative results for (a) foot ulcer bounding box detection on FUSeg dataset, (b) multi-class segmentation (\textcolor{lesion}{lesions}, \textcolor{skin}{skin}, and \textcolor{background}{background}) and in-the-wild body part prediction, (c) skin segmentation and body part prediction on Pratheepan dataset, and (d) multi-class segmentation (\textcolor{lesion}{lesions}, \textcolor{skin}{skin}, and \textcolor{background}{background}) on dermoscopy images from PH2 dataset. The color legend is the same as \figref{fig:AnnotationOverview}.
    }
    \label{fig:fitz17kResults}
\end{figure*}

\begin{table}[htb]
     \centering
     \caption{Foot ulcer bounding box detection and segmentation performance on the test set of real images of wounds.}
    \renewcommand{\arraystretch}{1.2}
    \resizebox{\columnwidth}{!}{
        \begin{tabular}{@{\extracolsep{5pt}}lcccc@{}}
    	    \toprule

                      & \multicolumn{2}{c}{Detection (bounding box overlap)} & \multicolumn{2}{c}{Segmentation (pixel-wise comparison)}\\
                      \cline{2-3}  \cline{4-5}
    		Train dataset  & $\mathrm{AP}_{\mathrm{centroid}}$ & IoU &  Dice & IoU\\
    		\midrule
    		Synthetic & 0.80  $\pm 0.018$          & 0.42 $\pm 0.011$  & 0.49  $\pm 0.007$           & 0.37 $\pm 0.008$\\
    		FUSeg   & 0.88  $\pm 0.012$          & 0.61  $\pm 0.008$  &   0.81 $\pm 0.003$           & 0.71 $\pm 0.004$ \\
    		\bottomrule
    	\end{tabular}
    }
	\label{table:fuseg-det-and-seg}
\end{table}

\subsection{Lesions, Skin, and Background Segmentation Using \emph{in-the-wild} Clinical Images}
\label{sec:segmentwild}

For our subsequent experiments, we modify the DeepLabV3 ResNet50~\cite{Chen2017} CNN model to perform two semantic segmentation tasks: skin condition vs healthy skin vs non-skin segmentation; and anatomical semantic segmentation (\secref{sec:anatomy}).
We add a total of 11 output channels for semantic segmentation, where three channels are used to predict the pixels containing skin conditions, healthy skin, and non-skin regions, and the remaining eight channels predict the anatomical labels.
Rather than use the full 16 anatomical labels provided as per SCAPE body model~\cite{scape2005}, we follow the PASCAL-part convention~\cite{chen2014detect} similar to~\cite{parsing_tested} and group the semantically similar anatomical labels into a single label (e.g., ``left upper arm" and ``right lower arm" are given the label ``arm") as shown in \figref{fig:3D_body_part_assignment}-(g).

We evaluate our approach on our manually annotated images taken from
Fitzpatrick17K, an ``in-the-wild" clinical 2D image dataset, where we use a subset of 25 images that were neither used for blending nor during model validation.
We use our CNN model trained on synthetic data and evaluate the performance on these real images.
We tested on these 25 manually segmented images and calculated the per-image Jaccard index for skin condition, skin, and non-skin segmentation.
The averaged results were $0.61 \pm 0.23$, $0.88 \pm 0.10$, and $0.60 \pm 0.43$, respectively.
We show the qualitative results in \figref{fig:fitz17kResults}.
These results suggest that the model is capable of generalizing from our synthetic data to real images.

\subsection{Lesion and Skin Segmentation on Dermoscopy, Clinical, and Non-Medical Images}
\label{sec:results:segment:other}
To further show the generalization capability of our approach, we use the same trained model described in \secref{sec:segmentwild} and evaluate over {PH2}~\cite{Mendonca2013,Ferreira2012}, a dermoscopy dataset with 200 images; {DermoFit}~\cite{Ballerini2013}, a clinical dataset of 1300 images; and the {Pratheepan}~\cite{Yogarajah2010,Tan2012} non-medical image dataset that provides manually segmented skin masks.
Interestingly, we find that our model, trained on synthetic data simulating ``in-the-wild" clinical images, does show the ability to generalize to these other types of datasets.
When segmenting lesions from the dermoscopy PH2 dataset, we achieve a Jaccard index of $0.62 \pm 0.21$ averaged over each image.
For context, previous work by~\cite{abhishek2020illumination} noted that applying transfer learning from curated real clinical images of close-up views achieves slightly better performance (Jaccard index of $0.69$).
When segmenting lesions from the clinical DermoFit dataset, we achieve a Jaccard index of $0.57 \pm 0.21$.
On the non-medical Pratheepan dataset, we use the 32 images in the ``FacePhoto'' partition, which shows a single closeup of a human subject, and achieve a Jaccard index of $0.76 \pm 0.14$.
For context, we report an F1-score of 0.86 while prior work~\cite{Buza2017} reported $0.74$.
While these do not represent state-of-the-art results on these datasets, we \emph{emphasize} that the purpose of these experiments is to show that our proposed approach to generate synthetic data results is of sufficient quality that a model can learn to generalize to real images.
We highlight that this single model trained on the synthetic labels predicts dense semantic segmentations for both the skin task and the anatomy task, and show the qualitative results of our model predicting other types of tasks in \figref{fig:fitz17kResults} (b)-(d).

\subsection{Predicting Body Parts from 2D Images}
\label{sec:results:predict:anatomy}
\begin{figure}[htb]
    \centering
    \includegraphics[width=\linewidth]{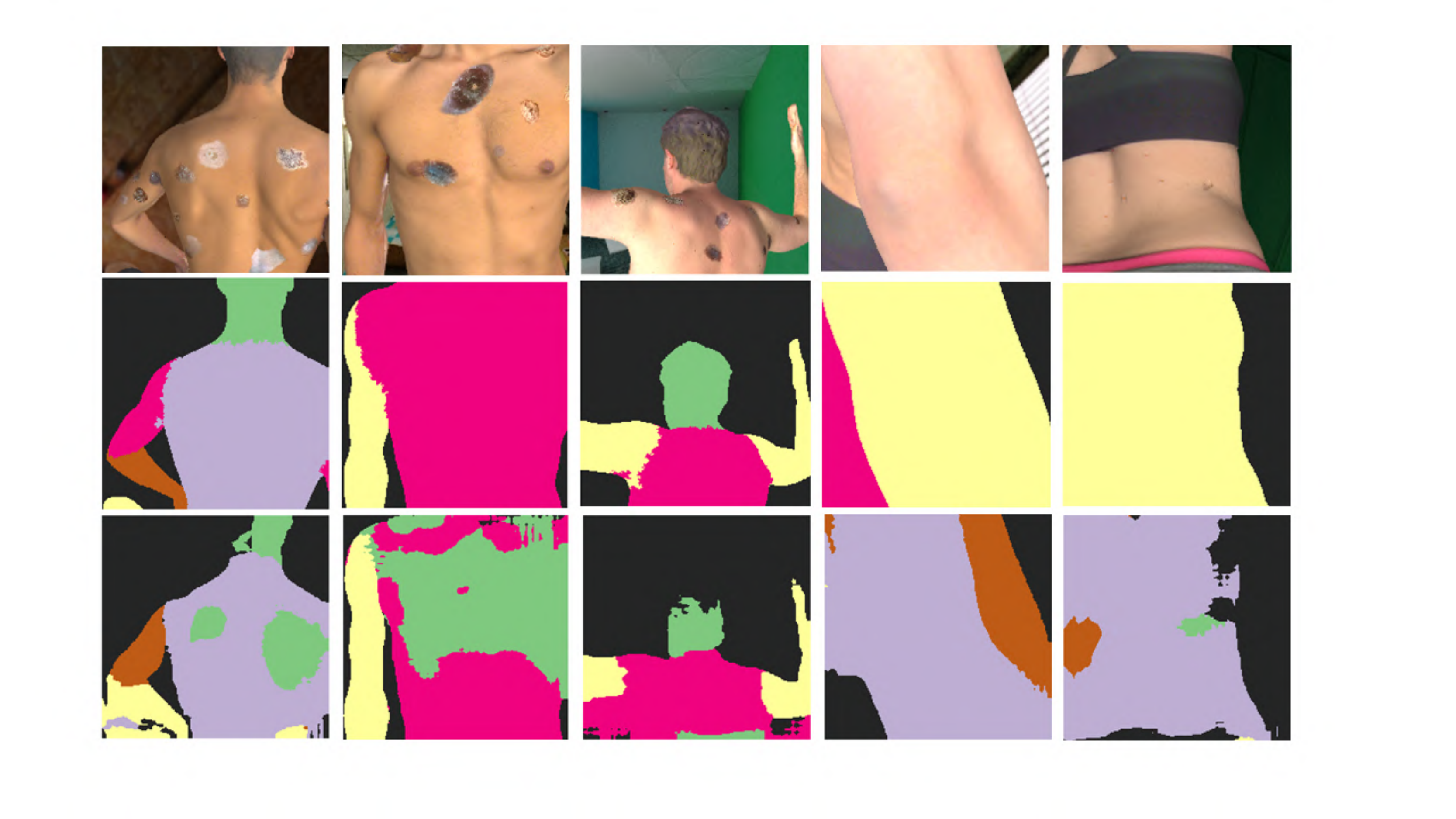}

    \caption{Qualitative results obtained by applying an existing human parsing method~\cite{parsing_tested} on the proposed dataset. Top: RGB images. Middle: ground-truth anatomical labels. Bottom: predicted anatomical labels.}
    \label{fig:anatomy_results}
\end{figure}

\begin{figure*}[hbt]
     \centering
     \begin{subfigure}[b]{0.49\textwidth}
         \centering
         \includegraphics[width=0.45\textwidth]{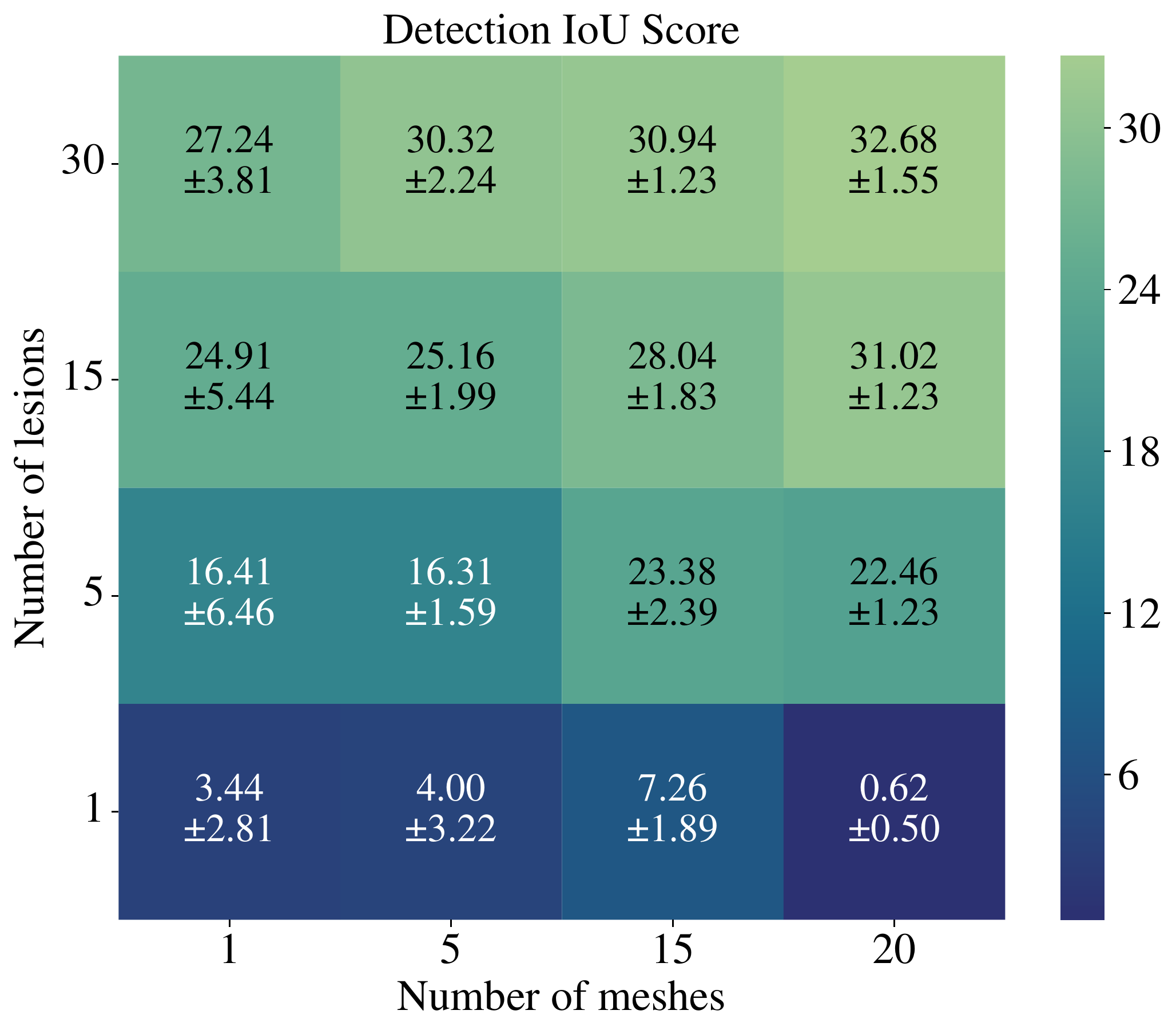}
         \includegraphics[width=0.45\textwidth]{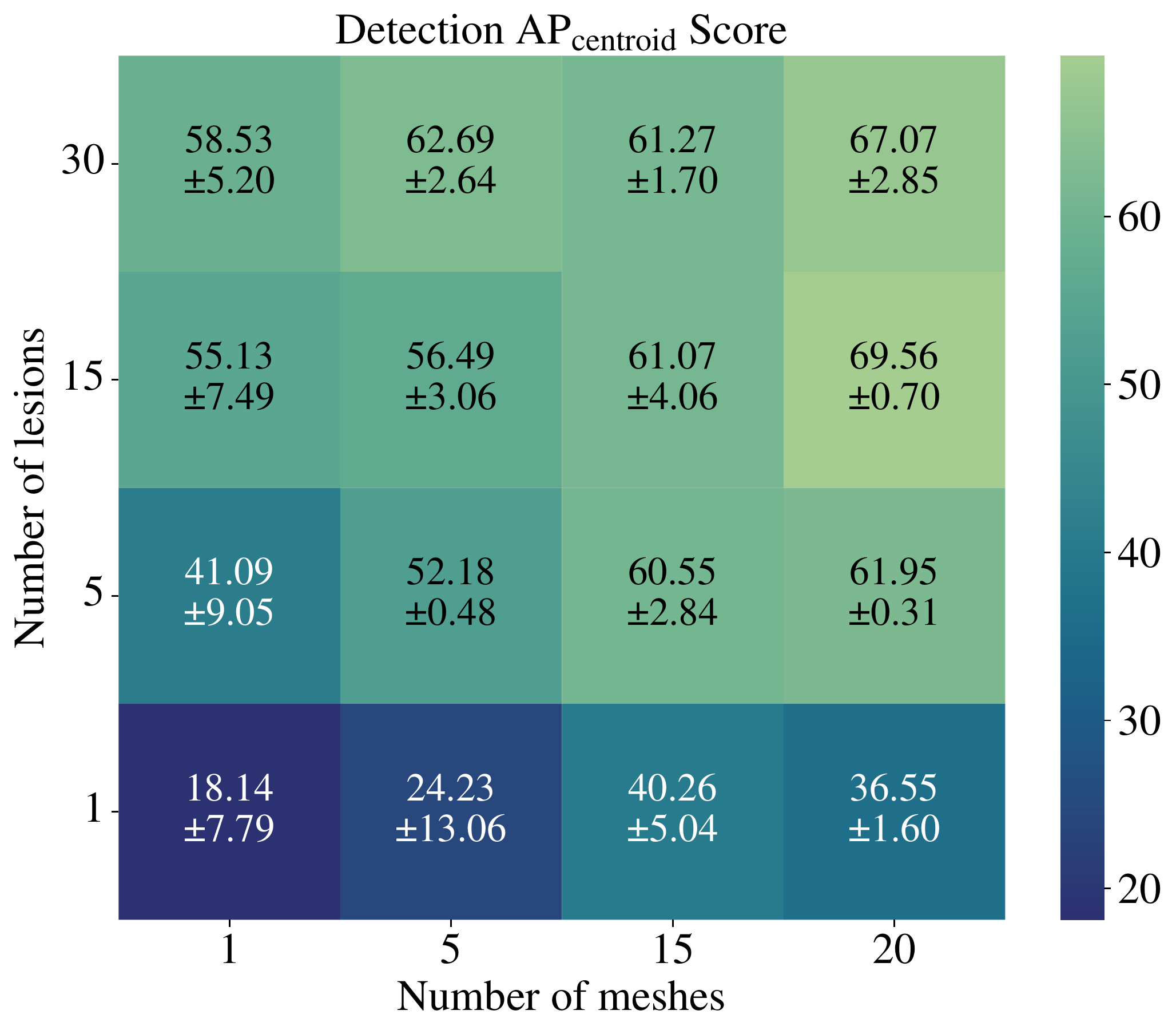}
         \caption{Foot Ulcer Detection}
     \end{subfigure}
     \begin{subfigure}[b]{0.49\textwidth}
         \centering
         \includegraphics[width=0.45\textwidth]{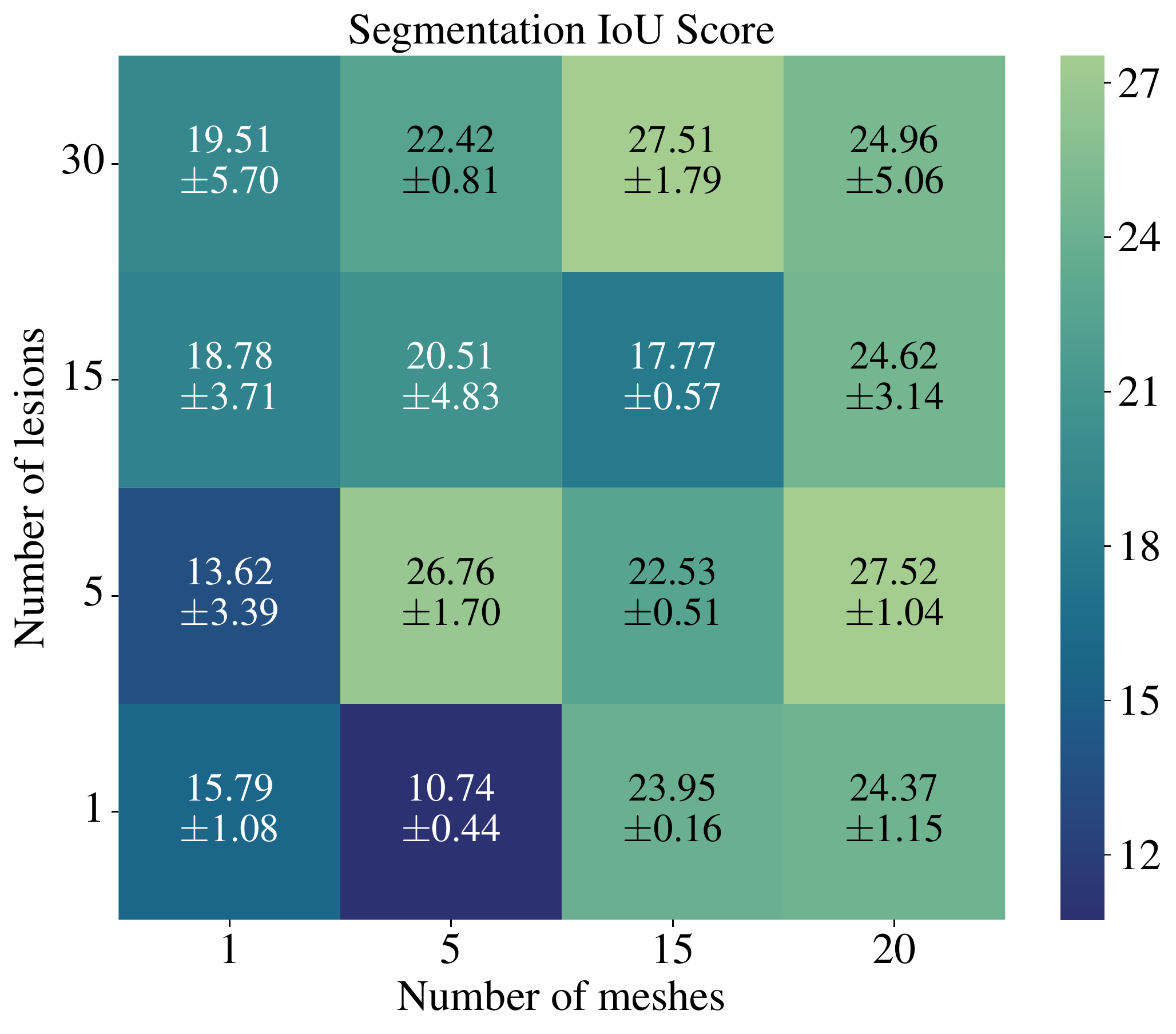}
         \includegraphics[width=0.45\textwidth]{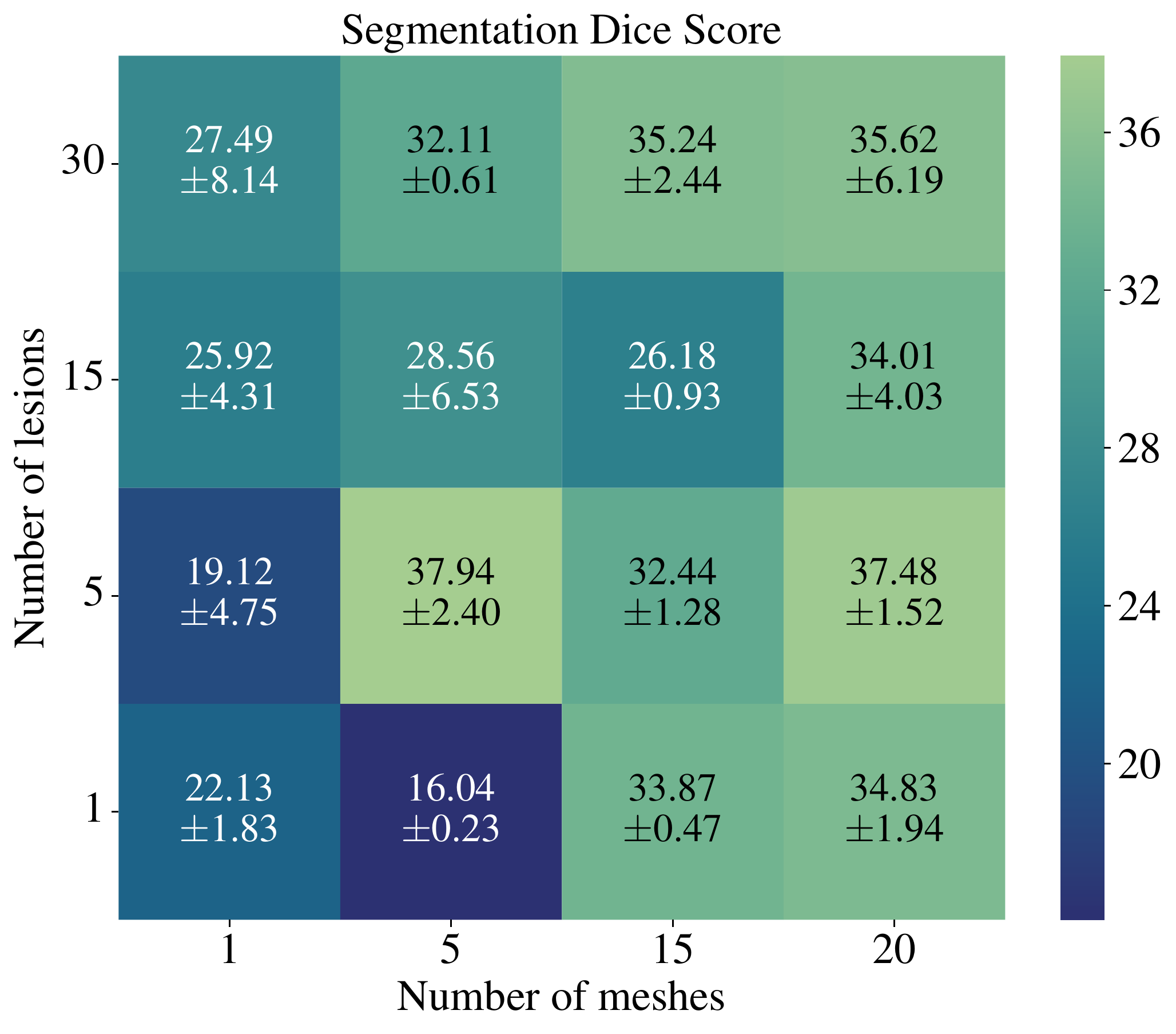}

         \caption{Foot Ulcer Segmentation}
     \end{subfigure}
        \caption{\ash{
        Ablation study on the effect of number of lesions and number of meshes on the downstream tasks of binary segmentation and bounding-box detection is visualized as a heatmap. The darker the shade, the lower is the value of the performance metric. We observe that the performance generally improves with an increase in the number of lesions and meshes, \ie, subjects, owing to the increase in diversity of the synthetic images.}
        }
        \label{fig:ablation}
\end{figure*}
Understanding where the skin condition is on the body may be an important factor in determining likely skin conditions~\cite{crombie1981distribution,pearl1986anatomical,youl2011body} (e.g., ruling out a diagnosis of a foot ulcer if the condition appears on the torso).
While semantic segmentation of the human body, commonly referred to as human parsing, is a well-studied area, we are specifically interested in partial-body views.
Existing approaches~\cite{parsing2,parsing_tested,parsing1} mainly consider constrained scenarios where the human body is fully visible in the images, but performance drops considerably for partial views.
Extreme close-up views of the anatomical body parts are challenging to discern, even to a human observer.
Conversely, distant views of the body expose a significant portion of the anatomy, making it easier to identify the body parts.
Therefore, we expect a higher level of accuracy in predicting the body locations on closer views of human body.
This hypothesis is confirmed by evaluating the performance of a recent human parsing method~\cite{parsing_tested} on the partial human body views from \method{}.
\ash{
Furthermore, the potential ambiguity in labeling the anatomical structure from close-up views, as observed by~\cite{sitaru2023automatic}, may be mitigated by our method which can create anatomical labels by mapping the anatomy visible in the view to 3D avatars that are registered on a standard template model~\cite{scape2005} consisting of $16$ annotated body parts.
Therefore, while rendering, a precise determination of the depicted body part is facilitated by the established semantic correspondence between the avatar and the template model.
}
This synthetic data is composed of a wide variety of views, going from close-up to distant views.
Our experiments show that the method proposed by~\cite{parsing_tested} achieved a mean IoU of only $0.29 \pm 0.20$ when tested on 100 randomly selected close-up human body views from~\method{}, as compared to an IoU of $0.63$ on the Pascal-person-part dataset~\cite{chen2014detect}.
These quantitative results align with the qualitative results presented in \figref{fig:anatomy_results}.
Moreover, the drastic drop in performance, both qualitatively and quantitatively, when testing on close-up partial body views highlights the importance of developing human parsing approaches that can handle partial and extreme close-up views.

\section{\ash{Ablation Study}}
\label{sec:ablation}
\ari{
To investigate the effect of parameter choices of image synthesis on the end results, we performed an ablation study on one of the potential use cases of the proposed framework: foot ulcer image analysis.
As manual segmentation and the acquisition of skin lesions and textured meshes are arguably the most cost-intensive variable options in DermSynth3D image generation, we focus on assessing the effect of different numbers of lesions and meshes on the performance of foot ulcer bounding box detection and segmentation.
We gradually vary the number of lesions and blend them on different numbers of meshes.
Subsequently, we generate a training set of 1500 images by capturing randomly rendered views of the 3D meshes employing a similar approach as described in \secref{sub:sec:dataset_const}.
Following the experimental settings outlined in \secref{sub:sub:sec:wound_exp}, we evaluate the method using the real evaluation set of the FUSeg dataset, and report the results in \figref{fig:ablation} using a heatmap.}

\ari{
 \figref{fig:ablation} shows the performance detecting (\figref{fig:ablation} a) and segmenting (\figref{fig:ablation} b) wounds when changing the number of lesions and meshes used to generate the synthetic data.
 Increasing the number of lesions and meshes used to generate synthetic images shows a general trend of improved performance.
 We attribute the variability within this trend to be partly due to the random sampling of meshes, lesions, and rendering parameters used to create the training datasets.
 We also observe that wound detection (using bounding boxes) exhibits more consistent improvements than wound segmentation (using dense pixel-wise predictions).
 This difference is likely because detecting lesion bounding boxes is more similar to detecting wound boundary boxes, than segmenting lesions is to segmenting wounds (where the precise pixel-wise boundary annotations of the wounds and lesions can differ).
 }


\ari{
We believe the overall segmentation performance can be further improved by either utilizing algorithms that attempt to reduce the domain gap between the synthetic and real data distributions or defining a training distribution that closely aligns with the test set.
}

\section{Conclusions}
\label{sec:conclusion}
We introduce \method{}, a novel framework
for synthesizing densely annotated \emph{in-the-wild} dermatological images by blending 2D skin conditions onto textured 3D meshes of human subjects using a differentiable renderer and generating a custom dataset of 2D views with corresponding labels that span across several downstream tasks, such as segmentation and detection.
Our results show the effectiveness of the generated synthetic data for selected dermatological applications, as demonstrated by the generalization achieved after training on synthetic data and testing on real data.
However, there are some limitations of our approach, including the design choices in the different steps, such as the sub-optimal selection of lighting parameters and camera positions, and a blending loss that may not preserve the diagnostic quality or accurately match the scale and the skin tone of the lesions.
Additionally, we only blended skin conditions that could be confidently manually segmented, and hence, did not include diffused skin disease patterns such as acne.
Despite these limitations, our results suggest that \method{} has the potential to generate meaningful dermatological data for computerized skin image analysis, especially in resource-constrained or ethically challenging real-world scenarios.
By open-sourcing our framework, we enable the research community to investigate various rendering settings such as different textured meshes, lighting and material properties, and blended skin conditions.
Furthermore, researchers can utilize our framework to extend the proposed methodologies and tackle other downstream tasks.
For instance, domain adaptation methods can be utilized to improve the segmentation and detection performance on real data (\figref{fig:synth_augment}) by leveraging the generated synthetic data.
\ash{
Exploring the performance of~\cite{sitaru2023automatic} on data produced by \method{} and analyzing the tradeoff between data collection, annotation burden and performance accuracy across varying proportions of real and synthetic data would be an intriguing avenue for investigation.
Moreover, diffusion-based modelling~\cite{ho2020denoising} may be a promising alternative to the current blending approach to achieve photorealism and generate diverse images while adhering to the same disease-class~\cite{gal2022textual}.
}

\section*{Acknowledgments}
The authors are grateful to Megan Andrews and Colin Li for their assistance with the data annotation efforts, which included the manual segmentations of non-skin regions in the texture images.
This research was enabled in part by support provided by the Natural Sciences and Engineering Research Council of Canada (NSERC), the Collaborative Health Research Projects (CHRP) program, the BC Cancer Foundation, and the BrainCare BC Fund, and the computational resources provided by WestGrid (Cedar),  \href{alliancecan.ca}{Digital Research Alliance of Canada} and NVIDIA Corporation.

{\small
\bibliographystyle{ieee_fullname}
\bibliography{main}
}

\clearpage{}%
\appendix

\section{Criteria for Skin Condition Location}
\label{sup:sec:methods}

We provide supplementary details related to \secref{sec:blend_locations}, which describe the criteria used to choose where on the mesh to blend a skin condition. We dilate the segmented skin condition \gls{segMask} and apply the same procedure in \secref{sec:blend_locations} to form an image of the dilated skin condition $a_{x_d}$ and its corresponding dilated mask $a_{s_d}$, which has an enlarged boundary to include pixels on the outside of the original mask boundaries. We check if the region within the dilated mask is suitable for blending by following the criteria outlined in \secref{sec:blend_locations}, which we include here for clarity: the region (1) should have minimal depth changes to help prevent blending lesions across disjoint anatomy; (2) should not overlap with the background; and, (3) should not overlap with clothes or the hair on the head. When blending multiple skin conditions, we also ensure that skin conditions do not overlap.

For the first and the second criteria, we get the depth \gls{viewDepth2d} from the renderer (\eqnref{eq:renderer}), where positive values indicate the distance from the mesh to the camera and negative values indicate pixels outside the mesh. By setting positive values to 1 and negative values to 0, we determine a mask for the body $a_{\mathrm{body}}$. The third criterion requires us to distinguish between skin and non-skin regions (e.g., clothing). For a texture image, we manually annotate a non-skin texture binary mask $T_{\mathrm{nonskin}}$,
the annotation process for which is described in \secref{sec:3dbodytex}.
We create a skin mask for the view by using $T_{\mathrm{nonskin}}$ as the texture image in \eqnref{eq:renderer} and rendering the view to create a binary mask of the non-skin regions \gls{viewNonSkinMask2d}. We combine the non-skin mask with the body mask $a_{\mathrm{body}}$ to compute a skin mask, $a_{\mathrm{skin}} = a_{\mathrm{body}} \odot (1-\gls{viewNonSkinMask2d})$. This skin mask is used to mask out the depth regions that occur on non-skin regions, $z_{\mathrm{skin}} = a_{\mathrm{skin}} \odot \gls{viewDepth2d} + (a_{\mathrm{skin}} - 1)$. We compute the maximum change of depth within the skin condition dilated mask $a_{x_d}$,
\begin{equation}
    c = \mathrm{max}(|z_{\mathrm{skin}} - \gls{weightDistance}| \odot a_{x_d} )
\label{eq:depthChange}
\end{equation}
where we compute the absolute difference between the depth of the skin pixels and the scalar weight \gls{weightDistance} from \eqnref{eq:camera_position}, which is the distance between the camera and the selected face. The returned scalar $c$ will be high when the skin condition overlaps with the background, non-skin region, or is spread across anatomy with a large change of depth, and $c$ will be low when the skin condition is on a skin region that is relatively flat with respect to the camera position. If $c$ exceeds a user-supplied threshold (which is a dataset-dependent value that we empirically set to $0.02$), we reject the view and sample a different face.

\section{Materials: Datasets and Annotations}
\label{sup:sec:materials}

\begin{figure*}[htb]
    \centering
    \includegraphics[width=\linewidth]{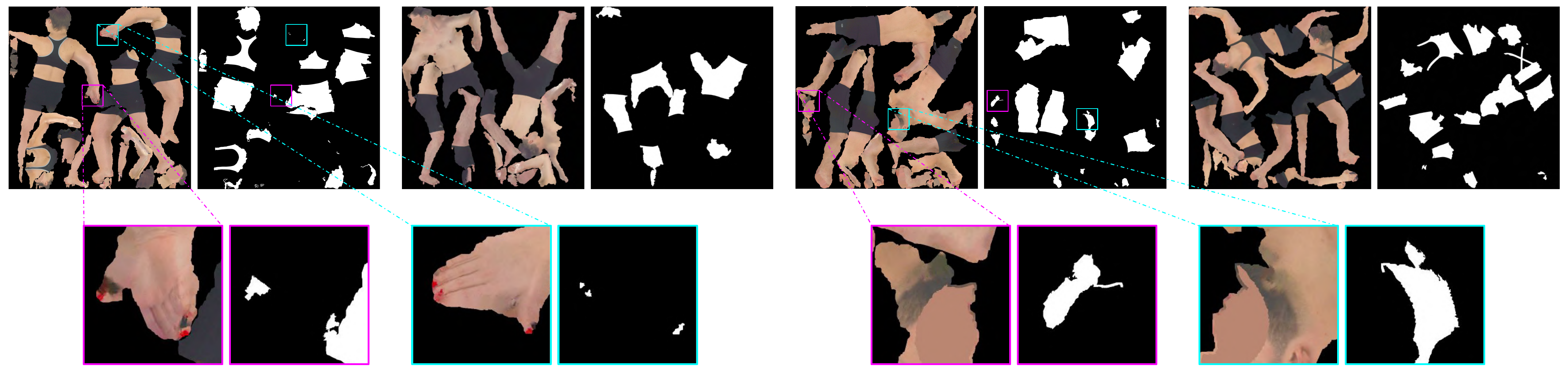}
    \caption{Samples from 3DBodyTex showing annotations of non-skin regions. For each pair, the left image shows the texture image, while the right image shows the corresponding non-skin region mask. The texture images are densely annotated to also exclude non-skin regions such as nail polish (first image) and facial beard (third image).}
    \label{fig:nonskin}
\end{figure*}

\begin{figure*}[htb]
    \centering
    \includegraphics[width=\linewidth]{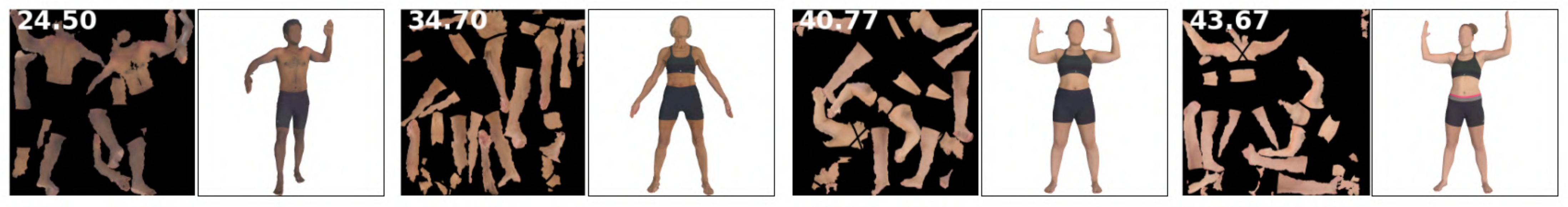}
    \caption{Samples from 3DBodyTex showing a range of skin-tones. For each pair, the left image shows the texture image with the ITA value on top left corner, while the right image shows a 2D view of mesh.}
    \label{fig:ita_skin}
\end{figure*}

\begin{figure}[hbt]
    \centering
    \includegraphics[width=\linewidth]{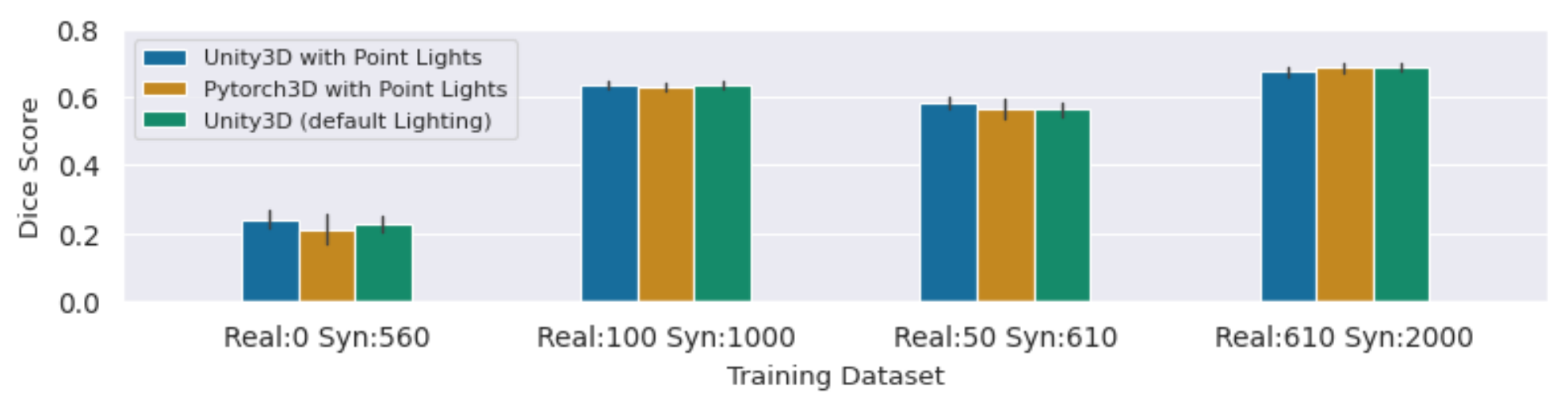}
    \caption{An ablation study on the choice of renderer for generating the 2D images, showing a performance comparison for wound segmentation task, on the test set of FUSeg dataset. The Y-axis represents the mean Dice score on the real ``test set" over 3 folds, and the X-axis represents the ``training set" which is a mix of generated synthetic data and real samples from FUSeg dataset. Each three-colored bar on the X-axis denotes the type of renderer used for generating the synthetic \method{} dataset, namely {\textcolor{cb2}{Pytorch3D with PointLights}, \textcolor{cb3}{Unity3D (default lights)}, and \textcolor{cb1}{Unity3D with Point Lights}}.}
    \label{fig:renderSeg}
\end{figure}

\subsection{Annotating Non-skin Regions}
\label{sup:sec:annotate_nonskin}

We provide supplementary details related to \secref{sec:3dbodytex}, which describe our approach to manually annotate 3DBodyTex~\cite{Saint2018,Saint2019} texture images. To annotate non-skin regions within the texture images, we used a semi-automated approach that selects contiguous regions based on color and user-supplied seeds and color thresholds, followed by manual free-hand correction where necessary.
We used the image editing software GIMP~\cite{gimp} to annotate a total of $168$ texture images, samples from which are shown in Figure~\ref{fig:nonskin}.
We selected a subset of 50 meshes to perform blending, where meshes were selected in order to sample from a range of skin tones available within 3DBodyTex.
Following prior works~\cite{Groh2021}, we estimate the range of skin tones as shown in Figure~\ref{fig:ita_skin} by computing the individual typology angle (ITA) over the skin regions and excluding the non-skin regions \gls{nonSkinTextureImage}.

As 3DBodyTex contains real human subjects, the texture images can contain real lesions. We leverage the manual bounding box annotations provided by~\cite{zhao2021skin3d} that localize existing pigmented skin lesions (e.g., a mole) on the 3DBodyTex texture images. We encode these bounding boxes as binary masks that correspond to the dimensions of the texture images, which are then used during rendering and incorporated into our synthetic labels.

\begin{figure}[htb]
    \centering
    \includegraphics[width=\linewidth]{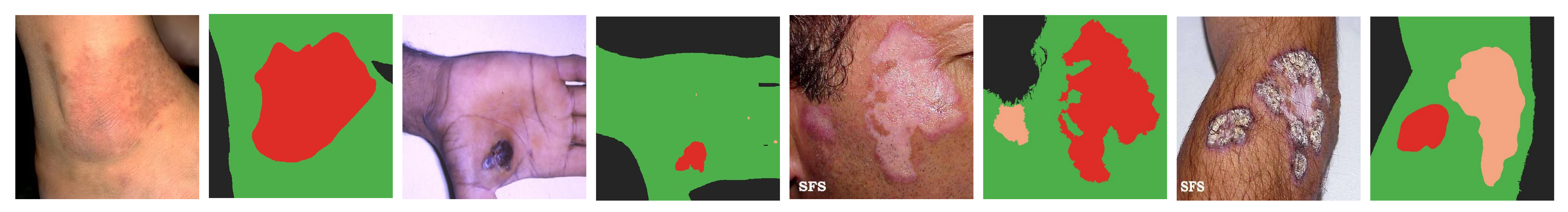}
    \caption{Dense annotations of skin lesions from Fitzpatrick17k. The annotation labels are: the selected lesion for blending (red), other lesions (brown), healthy skin (green) and background (black).}
    \label{fig:manual_masks}
\end{figure}

\subsection{Manual Segmentation of Fitzpatrick17k Images}
\label{sup:sec:annotate_fitz}

We provide supplementary details related to \secref{sec:fit17k}, which describe our process to manually annotate skin conditions within Fitzpatrick17K images~\cite{Groh2021}. When manually segmenting these 2D clinical images, we observed that for many of the lesions, it is challenging to determine the precise lesion boundaries. This is further complicated by some lesions exhibiting diffuse patterns, where many small lesions occupy a fraction of the image (e.g., acne), or the diseased regions only differ from the surrounding healthy skin based on the skin pigmentation (e.g., vitiligo). Thus, to reduce ambiguities in our manual segmentations, we perform an initial step to manually select images with a lesion that occupies a relatively large fraction of the image and exhibits well-defined borders. This limits the types of lesions we blend and qualitatively evaluate, and is a limitation of our work.

Using this subset of Fitzpatrick17K images, we manually segment lesions for blending into the texture image (Section~\ref{sec:blending}). When choosing lesions to segment for blending, we defined the following guidelines: (1) the lesion should be entirely visible within the image to avoid blending a lesion with an abrupt boundary; (2) the lesion should not be against the border of the image as we consider the surrounding skin during the blending; and (3) the lesion should be located on a relatively flat region of the body with minimal changes to the underlying anatomy (e.g., avoid choosing lesions in areas such as the armpit which contains both geometry changes and changes to the underlying skin texture that may not be specific to the lesion characteristics). This will prevent distortions and blending of the features of the underlying anatomy with the characteristics of the lesion.

\begin{table*}[htb]
\centering
\caption{Metadata statistics for the segmented lesions from Fitzpatrick17k~\cite{Groh2021}.}
\label{tab:Fitzpatrick17k_stats}
\begin{tabular}{ccccccccccc}
\toprule
\multirow{2}{*}{\textbf{Split}} & \multirow{2}{*}{\textbf{\begin{tabular}[c]{@{}c@{}}\# images\\ (\# diagnosis labels)\end{tabular}}} & \multicolumn{3}{c}{\textbf{Three-partition Label}}                                                      & \multicolumn{6}{c}{\textbf{Fitzpatrick Skin Type}}                                                                                                                                  \\ %
                                &                                                                                                     & \multicolumn{1}{c}{\textbf{benign}} & \multicolumn{1}{c}{\textbf{malignant}} & \textbf{non-neoplastic} & \multicolumn{1}{c}{\textbf{1}} & \multicolumn{1}{c}{\textbf{2}} & \multicolumn{1}{c}{\textbf{3}} & \multicolumn{1}{c}{\textbf{4}} & \multicolumn{1}{c}{\textbf{5}} & \textbf{6} \\
                                \midrule
Validation                      & 50 (30)                                                                                             & \multicolumn{1}{c}{11}              & \multicolumn{1}{c}{17}                 & 22                      & \multicolumn{1}{c}{5}          & \multicolumn{1}{c}{17}         & \multicolumn{1}{c}{15}         & \multicolumn{1}{c}{6}          & \multicolumn{1}{c}{6}          & 1          \\ %
Testing                         & 25 (21)                                                                                             & \multicolumn{1}{c}{6}               & \multicolumn{1}{c}{9}                  & 10                      & \multicolumn{1}{c}{6}          & \multicolumn{1}{c}{6}          & \multicolumn{1}{c}{6}          & \multicolumn{1}{c}{5}          & \multicolumn{1}{c}{2}          & 0          \\
\bottomrule
\end{tabular}%
\end{table*}

We also perform a dense segmentation of the selected subset of Fitzpatrick17K images in order to create validation and test data. For this task, we manually partition the image into lesion and non-skin regions, where the skin region can be inferred by the absence of either of these two regions. We define non-skin regions as those regions that do not include skin (e.g., clothing, hair that occludes the skin, background). We segment 50 and 25 lesion images for the validation and the test data respectively, and the high-level statistics for the metadata of these lesions are provided in \tabref{tab:Fitzpatrick17k_stats}. It is worth noting that despite the relatively small number of images, we capture a large diversity in terms of Fitzpatrick skin tones (all 6 skin tones for validation and 5 skin tones for testing) and diagnoses (30 disease labels for validation and 21 for testing).

To perform the segmentation, we use the same software and process as described in \ref{sup:sec:annotate_nonskin}. For each image, the resulting manual segmentations (Figure~\ref{fig:manual_masks}) are represented by the following binary masks: all lesions belonging to the corresponding disease type of the image $m_{\mathrm{lesions}}$, a selected lesion region for blending $m_{\mathrm{blend}}$, where $m_{\mathrm{blend}} \in m_{\mathrm{lesions}}$, and non-skin regions $m_{\mathrm{nonskin}}$. We infer a skin mask that excludes the lesion regions as $m_{\mathrm{skin}} = (1-m_{\mathrm{lesions}}) \odot (1-m_{\mathrm{nonskin}})$, where $\odot$ indicates an element-wise product.

\section{Experiments and Results}
\label{sup:sec:experiments}
\subsection{Wound Detection in Clinical Images}
\label{sup:sec:fuseg_det_seg}

We provide supplementary analysis related to the wound bounding box detection results in  \secref{sec:blend_locations}. We can see in Table \ref{table:fuseg-det-and-seg} that the model trained on only synthetic images achieves an $\mathrm{AP}_{\mathrm{centroid}}$ of $0.80$ and IoU of $0.42$.
The significant gap between the IoU and $\mathrm{AP}_{\mathrm{centroid}}$ suggests that the model localizes the wounds, but does not precisely match the bounding boxes encapsulating them.
By analyzing the qualitative results of the model's predictions (Figure \ref{fig:fitz17kResults} (a)), we observed two major trends in the model's failure cases. (1) There seems to be a semantic difference between a skin condition and a wound. In our synthetic dataset, the whole lesion area, including the surrounding affected skin, is annotated as the lesion. However, in the FUSeg dataset, only the open-wound area is covered by the segmentation mask. This mismatch in labeling across these two image domains causes the model to over-segment some images (Figure \ref{fig:fitz17kResults}(a) bottom three rows), resulting in a drop in the IoU. (2) As the synthetic data contains a variety of skin conditions across different parts of the body when trained on synthetic images, the model learns to detect other skin conditions within the image that are not of the wound. This can cause the model to over-detect wounds in the images (Figure \ref{fig:fitz17kResults} (a) bottom row), resulting in a decrease in both IoU and $\mathrm{AP}_{\mathrm{centroid}}$.

\subsection{Lesions, Skin, and Background Segmentation Using in-the-wild Clinical Images}
\label{sup:sec:lesion_skin_bg}

We provide further details on the model trained to predict semantic segmentations related to the skin and the anatomy, as described in \secref{sec:segmentwild}. We apply a softmax function computed over each spatial location across the skin condition, skin, and non-skin output channels, and a softmax function computed over each spatial location across the anatomical output channels. We modify the fuzzy Jaccard index to act as a loss and to ignore empty ground truth channels when computing the loss. For the anatomy channels, the Jaccard loss is computed over an entire channel within a batch, while for the other channels, the Jaccard loss is computed separately for each channel and each image. These modifications were made to address the different types of class imbalances that occur within the two different semantic segmentation tasks. In addition, when computing the loss over the skin channel, we ignore locations that contain the skin condition to reduce the impact of misclassifying healthy skin as the skin condition. For both the skin condition and skin channel, when computing the loss we ignore locations that overlapped with manually curated bounding boxes that signify the location of an existing mole on the original texture images (as determined by~\cite{zhao2021skin3d} and described in \ref{sup:sec:annotate_nonskin}) as the regions within the bounding box can contain both healthy skin and a lesion.

\begin{figure*}[htb]
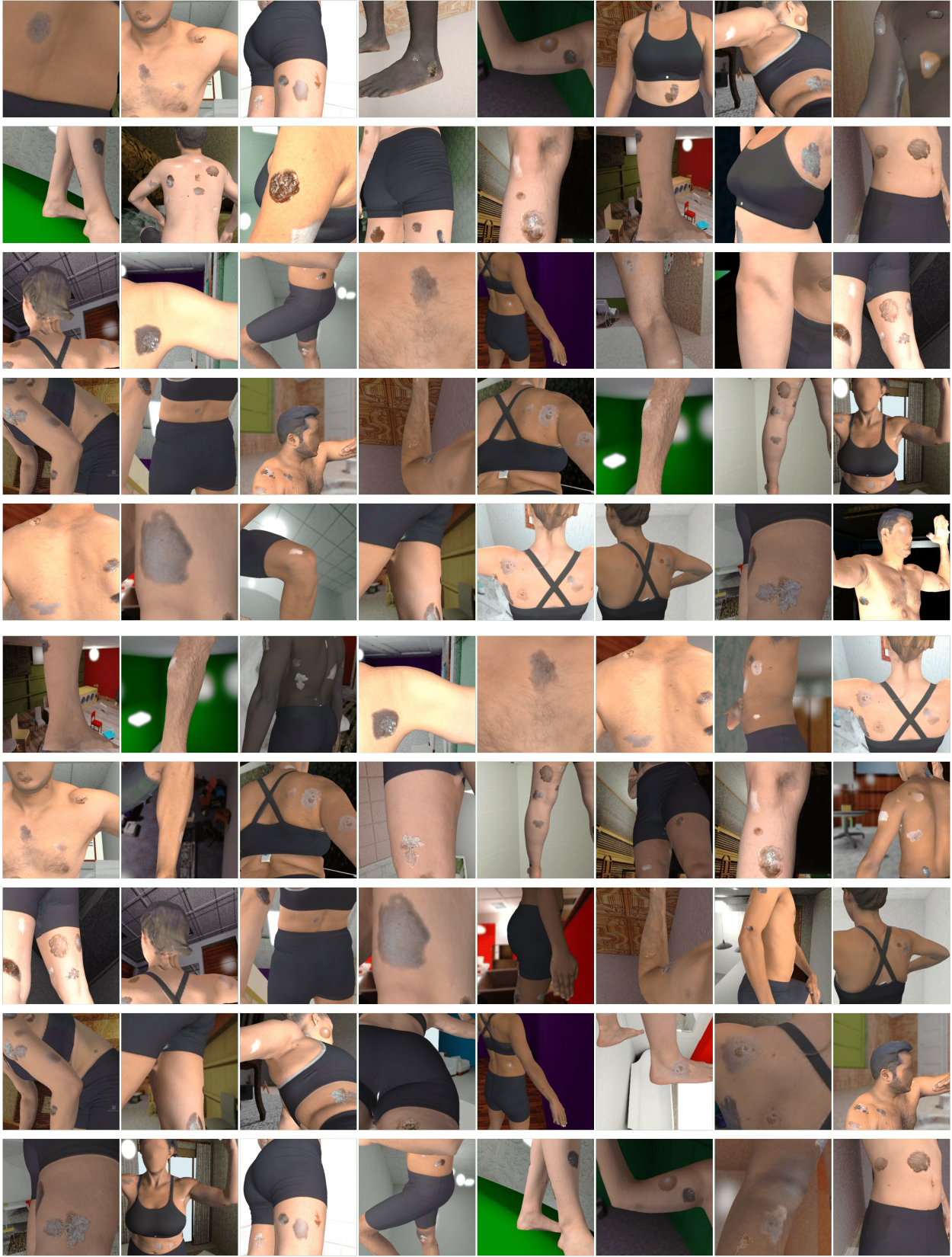

    \centering
    \begin{subfigure}[b]{\linewidth}
     \centering
    \includegraphics[width=0.95\linewidth]{fig_1-min.pdf}
    \end{subfigure}

    \begin{subfigure}[b]{\linewidth}
     \centering
    \includegraphics[width=0.95\linewidth]{fig_2-min.pdf}
    \end{subfigure}

    \caption[]{Additional samples of generated synthetic images of multiple subjects across a range of skin tones in various skin conditions, backgrounds, lighting, and viewpoints.
    }
    \label{fig:more_samples2}
\end{figure*}


\clearpage{}%
\end{document}